\documentclass[11pt]{article}
\usepackage[letterpaper, left=1in, top=1in, right=1in, bottom=1in,nohead,includefoot]{geometry}
\usepackage{color,charter,graphicx,here,setspace,multirow,amsmath,amssymb,amsfonts,enumitem,placeins,amsthm,cancel,mathrsfs,bm,pdflscape,subcaption}
\usepackage[utf8]{inputenc}
\usepackage[authoryear]{natbib}

\RequirePackage{amsthm,amsmath,amsfonts,amssymb}
\RequirePackage[authoryear]{natbib}
\RequirePackage{graphicx}
\usepackage{placeins} 
\usepackage{subcaption} 
\usepackage[all]{xy}

\usepackage[dvipsnames]{xcolor}
\usepackage[pdftex]{hyperref}
	\hypersetup{colorlinks,
	citecolor=NavyBlue  
}

\def\Eqno#1{Eqn.~(\ref{eq:#1})}\def\eqno#1{eqn.~(\ref{eq:#1})}

\def\Eqnos#1#2{Eqns.~(\ref{eq:#1}) and (\ref{eq:#2})}
\def\seq#1#2{#1{:}#2}\def\ci{\perp\!\!\!\perp}

\def\dj{\delta_{x_j}}\def\djb{\delta_{x_j-\beta_j}}\def\jj{{j=\seq1J}}

\def\mM{\mathcal{M}}\def\mD{\mathcal{D}}\def\mH{\mathcal{H}}

\def\bb{\mathbf{b}}\def\q{\mathbf{q}}\def\S{\mathbf{S}}\def\x{\mathbf{x}}\def\xj{{\mathbf{x}_{-j}}}
\def\m{\mathbf{m}}\def\u{\mathbf{u}}

\def\bbeta{\mbox{\boldmath$\beta$}}\def\bbetaj{\mbox{\boldmath$\beta_{-j}$}}\def\bgamma{\mbox{\boldmath$\gamma$}}
\def\bmu{\mbox{\boldmath$\mu$}}\def\muj{\mbox{\boldmath$\mu_{-j}$}}
\def\bSigma{\mbox{\boldmath$\Sigma$}}\def\btheta{\mbox{\boldmath$\theta$}}

\graphicspath{{figures/}}

\date{\today}

\begin{document}

\title{Bayesian Predictive Synthesis with Outcome-Dependent Pools}
\author{Matthew C. Johnson\thanks{Amazon, Seattle, WA 98109-5210 {\scriptsize{}Email: mcjohnson946@gmail.com}}\ \ \& Mike West\thanks{Department of Statistical Science, Duke University. Durham NC 27708-0251. {\scriptsize{}Email: Mike.West@duke.edu}}}

\maketitle


\begin{abstract}
This paper reviews  background  and examples of Bayesian predictive synthesis (BPS), and develops details in a subset of BPS mixture models. 
BPS expands on standard Bayesian model uncertainty analysis for model mixing to  
provide a broader foundation for calibrating and combining  predictive densities from multiple models or other sources. 
One main focus here is BPS as a framework for justifying and understanding generalized \lq\lq linear opinion pools,'' where multiple predictive densities are combined with flexible mixing weights that depend on the forecast outcome itself-- i.e., the setting of outcome-dependent model mixing.  BPS also defines approaches to incorporating and exploiting dependencies across models defining forecasts, and 
to formally addressing the problem of model set incompleteness  within the  subjective Bayesian framework.  In addition to an overview of 
general mixture-based BPS,  new methodological developments for dynamic BPS-- involving calibration and pooling of sets of predictive distributions in a univariate time series setting-- are presented.   These developments are exemplified in summaries of an analysis in a univariate financial time series study.

\bigskip{}
\end{abstract}
\newpage{}

\section{Introduction}
\label{sec:intro}

The combination of forecast densities,
whether they result from a set of models, a group of consulted experts, or other sources, continues to be an active and important research arena that cuts across a range of disciplines.
Requiring methodology that goes beyond standard Bayesian model uncertainty and model mixing-- with its well-known limitations based on a clearly proscribed theoretical basis-- multiple   density combination methods have been proposed.
In recent years, the literature has been particularly rich in development of density forecast combination methods motivated by applications in economics, policy, and finance \citep[e.g.][]{Amisano2007,HallMitchell2007,Hooger2010,Kascha2010,Geweke2011,Geweke2012, Billio2012,Billio2013,Aastveit2014,Fawcett2014,Pettenuzzo2016,McAlinnWest2018,West2020Akaike}. Other key areas of application are as  diverse as meteorology, military intelligence, seismic risk, and environmental risk, among others~\citep[e.g.][]{Clemen1989,ClemenWinkler1999,Timmermann2004,ClemenWinkler2007,rufo2012}. 
These and other ensemble, averaging or \lq\lq synthesis''  methods have varying goals, applicability, and degrees of applied success. While empirical results can be be positive and encourage interest in the underlying method, many such forecast combination \lq\lq rules'' lack any sort of generative model or foundational justification. This is a key point of departure and emphasis of Bayesian predictive synthesis (BPS), reviewed, explored and expanded upon here. 

The  literature on agent opinion synthesis, in which a decision maker solicits the opinions of experts in order to create an informed opinion, provides a framework for model combination.
Forecast synthesis fits naturally into the \lq\lq supra-Bayesian'' approach~\citep[e.g.][]{DVL1979,West1988,West1992c,West1992d}. Here, a single 
Bayesian decision maker regards the new information gained from a set of models- or \lq\lq agents''- based forecast distributions, and evaluates approaches to formally condition on this information to define  resulting predictions. 
Generalized density combination is approached within this framework.

Importantly and as discussed below, BPS enables the decision maker to address questions of model/agent specific biases and calibration,  dependencies across models/agents, as well as to explicitly address the issue of model set incompleteness \citep{AastveitEtAl2019,McAlinnEtAl2020,McAlinn2021,GiannoneEtAl2021}, also known as the ``model space open" or ``$\mM$-open" setting \citep{BernardoSmith1994,West1997,ClydeGeorge2004,ClydeIversen2013}.
These ideas are also relevant to the generalized Bayes literature and the question of how to combine densities admitting that all models are \lq\lq wrong but useful" \citep[e.g.,][]{deHeidi2019}.  

\paragraph*{Some Notation:} Vectors are denoted using lowercase bold font;  for example, 
$\bmu$ represents a vector while $\mu$ is a scalar. Matrices are in upper-case bold font. Index notation $\seq1J$   represents the sequence $1,2,\dots,J$. A column vector $(x_1,x_2,\dots,x_J)'$ is denoted $\x$ or, in context in terms of elements as $x_{1:J}$. The notation $\xj$
indicates the vector $\x$ or set $x_{\seq1J}$ with the  $j^\text{th}$ element omitted.   The usual notation $N(\bmu,\bSigma)$  denotes a normal distribution with mean vector $\bmu$ and covariance matrix $\bSigma$, with usage  $\x\sim N(\bmu,\bSigma)$ and $N(\x|\bmu,\bSigma)$ for the p.d.f.  
The Dirac delta function $\delta_x(y)$ is the point-mass at $x$ as a distribution for $y$.
Other specific notation is defined in context. 

\section{BPS Foundations}

\subsection{Background and Key Theory}
A decision maker $\mD$ is interested in predicting an uncertain quantity $y$. The decision maker has some opinion of $y$, quantified through a subjective prior density $p(y)$.
In order to predict $y$, $\mD$ will examine the density forecasts $h_j(y)$ from $J$ separate sources $\mM_j$, $j=\seq1J$.
In general, these sources could be models, analysts, other forecasters, or subject matter experts; here, refer to them as models throughout. 
How should $\mD$ consolidate this information, and ultimately update $p(y)$?

The Bayesian paradigm defines the straightforward solution, in theory. $\mD$ updates the prior to a posterior upon learning the information set $\mH=\{ h_1(\cdot),\ldots,h_J(\cdot) \}$.
Specifying a full prior joint distribution $p(y,\mH)$ is impractical, however, and the theoretically straightforward approach cannot be easily implemented. This led \cite{West1992d} and \cite{West1992c} to extend the   work of \cite{GenestSchervish1985} to show that, under certain consistency conditions, $\mD$'s posterior density has the form
\begin{equation}
    \label{eq:bps}
    p(y|\mH) = \int\alpha(y|\x)h(\x)d\x
\end{equation}
where $h(\x)=\prod_{j=\seq1J}h_j(x_j)$.
Here, $\x$ is a vector of \textit{latent model states}, and $\alpha(y|\x)$ is a conditional density function that synthesizes these states.
The framework allows flexibility in choosing the key {\em synthesis function}  $\alpha(y|\x)$.
One requirement is consistency with $\mD$'s prior, i.e., 
\begin{equation}
    \label{eq:consistency}
    p(y) = \int\alpha(y|\x)m(\x)d\x
\end{equation}
where $m(\x)=E[h(\x)]$ with the expectation taken with respect to $\mD$'s implicit prior over $\mH$. 

Operationally and constructively, this result can be noted to require that $\mD$ specify $m(\x)$ and the conditional density function $\alpha(y|\x)$, inducing the prior $p(y)$.  Specifying these two functions allows $\mD$ to incorporate views of the models in terms of past information and expectations on aspects of their calibration, biases, relative expertise, and importantly, dependencies, in predicting $y$.  There is a great deal of flexibility in the specification, and a number of approaches to forecast density combination can be represented this way~\citep{McAlinnEtAldiscussionBA2018}. Some examples are given below.

The Bayesian update of the prior in~\eqno{consistency} to the posterior in~\eqno{bps} simply substitutes $h(\x)$ for $m(\x)$. This is an example of formal subjective Bayesian updating via  {\em (Richard) Jeffrey's rule} rather than by Bayes' theorem.  Implicitly,  $m(\x)$ is $\mD$'s prior for the latent model states $\x$, while $h(\x)$ is their ``true'' distribution later observed. In this update, the conditional density $\alpha(y|\x)$ remains unchanged. 
This method of updating is implicit in all Bayesian analysis and was recognized and formalized in \cite{Jeffreys1990}; see further discussion in \cite{DiaconisZabell1982}.
\cite{Jeffreys1990} provides an intuitive example that involves $\mD$ betting on a racehorse that performs better in mud. The probability of the horse winning \emph{conditional on rain} is known, but the probability of rain is not.
$\mD$'s prior that the horse wins depends on $\mD$'s prior for rain; when a professional forecast for rain becomes available, this simply replaces $\mD$'s  prior for rain.

The product of the $h_j(\cdot)$ in the key expression in~\eqno{bps} {\em does not} reflect any assumption of independence across models.  Rather, it makes clear that the inherent latent factors $x_j$ are conditionally independent {\em given} $\mH.$  In contrast, the implicit, partially specified prior $\mD$ has over  $(y,\mH)$  allows for essentially arbitrary dependencies of the $h_j(\cdot)$ as uncertain functions.  This is an important conceptual point and a point of departure from traditional pooling methods including Bayesian model averaging (BMA).   For example, historical evidence of positive dependence between two models-- in terms of them having generated rather similar forecast distributions in the past-- can be reflected in the synthesis.   Of course, how this is done in any specific context depends on the form of the synthesis function $\alpha(y|\x)$ specified.  

The interpretation of $\alpha(y|\x)$ is a focus of discussion in \cite{West1992d} and \cite{West1992c}.  The first interpretation is that, if each model $\mM_j$ were to provide a predictive density degenerate at a point $x_j$, i.e.\ $h_j(y)=\delta_{x_j}(y)$, then $\mD$'s posterior is given by $p(y|\mH)=\alpha(y|\x)$.
A second interpretation is that in order to sample $y^*\sim p(y|\mH)$, $\mD$ may first sample a vector $\x^*$ from $h(\x)$, and then sample $y^*\sim\alpha(y|\x^*)$. 

Much of the review discussion to follow, and ensuing methodological developments of this paper, focus on the theoretical framework of BPS using specific discrete mixture forms for $\alpha(y|\x)$.   Among other things, this justifies  the approach termed {\em generalized linear pools} of \cite{Fawcett2014},  but then extends to address questions of model inter-dependencies and time-varying generalizations relevant to predictive synthesis in time series forecasting.  First, however, note a simple example that connects with other variants of BPS-- involving other choices of the synthesis functions $\alpha(y|\x)$--  and that provides an easily appreciated, entr\'{e}e  example as well as connections to the literature. 

\subsection{Linear Regression Example}
\cite{McAlinnWest2018} explore dynamic BPS examples in which the $\alpha(y|\x), m(\x)$ are extended to time-varying settings and are conditionally normal given relevant defining parameters.   A simple, static example involves 
a synthesis function $\alpha(y|\x)$ that is the p.d.f. of 
\begin{equation*}
    (y|\x,\theta,\upsilon)\sim N(\theta_0 + \btheta'\x, \upsilon).
\end{equation*}
This example, just one potential specification of the conditional synthesis density $\alpha(y|\x)$, easily and intuitively allows for ranges of model biases and miscalibration, viewed through shifts in
means and/or variances of implied conditional distributions of individual conditional distributions $(x_j|y)$, and for cross-model dependencies through the regression vector $\btheta.$ 
If any model (or all models) are biased, the intercept term $\theta_0$ allows a correction.
If models are correlated (or anti-correlated), appropriate adjustments are made through the $\theta_j$ weights.
Finally, the residual volatility $\upsilon$ accounts for the relative uncertainty between the outcome $y$ and the latent states $\x$ following
corrections for their biases and  dependencies. 

By observing repeated forecasts over time, Bayesian updating allows for learning about the $\theta_j$ coefficients, which account for evolving perceptions of model biases and dependencies.
This example has been extended to multivariate forecast density synthesis and applied in a detailed macroeconomic study \citep{McAlinnEtAl2020}. 
Related, more extensive developments of dynamic BPS underlie aspects of more recent studies in macroeconomics and allied areas~\citep[e.g.][]{McAlinn2021,AastveitEtAL2022}.

\section{Mixture BPS}
\label{sec:mixture}

\subsection{Setting and Background} 
Empirical methods that simply average a set of predictive densities to form a discrete mixture-- or \lq\lq linear pool''-- have a long history~\citep[e.g.][and references therein]{Clemen1989,GenestSchervish1985,ClemenWinkler2007,HallMitchell2007}. 
Averaging the $h_j(\x)$ with respect to defined weights, or  {\em mixture probabilities}, can arise more formally  from standard BMA~\citep{BernardoSmith1994,ClydeGeorge2004,ClydeIversen2013} and extensions in time-varying settings~\citep[e.g.][chapter 12]{West1997}. Allied approaches choose weights that aim to optimize defined predictive goals~\citep[e.g.][and references therein]{Geweke2011,
Diebold2019,Diebold2022,LavineLindonWest2021avs,LoaizaMayaJOE2021}. 

A significant expansion of the scope of linear density pooling was marked by \cite{Fawcett2014},  proposing the use of outcome-dependent weights. This involves the {\em generalized linear pool} form 
\begin{equation}
\label{eq:generalizedpool}
p(y|\mH)=\sum_{j=1:J}w_j(y)h_j(y)
\end{equation}
where the weights $w_j(y)$ explicitly depend on the as-yet unobserved outcome being predicted and are defined so that $p(y|\mH)$ is a p.d.f. 
The authors presented this as an empirical approach to extending the traditional linear opinion pool using constant weights.  The core idea is that, in some regions of the outcome space of $y$, different models $\mM_j$ may be expected to generate superior forecasts. For example, one model may be generally better at predicting changes in inflation when inflation is high, another superior in times when inflation is low or falling.  
With more flexible weight {\em functions} fitted using semi-parametric Bayesian methods,~\cite{Fawcett2014} give examples of substantial improvements in forecasting accuracy  over traditional linear pools.  This has been followed by various extensions and demonstrations of empirical success using
outcome-dependent density pooling~\citep[e.g.][]{Aastveit2014,Pettenuzzo2016,BassettiEtAl2018}. While these techniques make intuitive sense as a generalization of constant weight approaches, and have been shown to produce good empirical results in various applications, their use  raises questions about foundations and theoretical justifications that would aid in understanding, extensions and identification of limitations. These questions are addressed within the framework of BPS. 

\subsection{Constant Weight Mixture BPS \label{sec:constantmixBPS} }

Suppose $\mD$ adopts the synthesis function  
\begin{equation}
    \label{eq:constant_weights}
    \alpha(y|\x) = \omega_0h_0(y) + \sum_{j=1:J}\omega_j\delta_{x_j}(y)
\end{equation}
where the weights $\omega_{0:J}$ are non-negative probabilities that sum to one.
Here,  $h_0(y)$ is a {\em baseline p.d.f.},  possibly defined by a further {\em baseline model}, that is  chosen to represent a 
\lq\lq save haven'' predictive distribution that $\mD$  will choose to revert to in case models $\mM_{1:J}$ are regarded as suspect, to be down-weighted as a group, based on poor predictive performance. The choice of a rather diffuse baseline p.d.f. is natural and links to other areas of the Bayesian forecasting literature where \lq\lq diffuse alternatives'' are chosen for comparison with predictions from one or more models~\cite[e.g.][section 11.4]{West1997}.  The inclusion of this baseline model in the linear pool of~\eqno{constant_weights} opens the path to addressing the $\mM$-open, or model set incompleteness, question.  Otherwise,~\eqno{constant_weights} is 
inspired by the interpretation that, if each model provides an \lq\lq oracle'' prediction $y =x_j$, then $\mD$ will linearly average these values and combine with the baseline.  Of course, the $x_j$ are latent variables; ~\eqno{bps} recognizes that and results in  $\mD$'s predictive density
\begin{equation*}
    p(y|\mH) = \omega_0h_0(y) + \sum_{j=1:J}\omega_jh_j(y).
\end{equation*}
Note that choosing $\omega_0=0$ results in classic linear pools of model densities, including many that previously had mainly empirical justification. This also includes formal Bayesian approaches via BMA and its extensions; that is, traditional Bayesian model uncertainty and combination is a special case of BPS. 
 
Evaluation of \eqno{consistency} provides interesting insight.  $\mD$'s prior (before observing $\mH$) is
\begin{equation*}
    p(y) = \omega_0h_0(y) + \sum_{j=1:J}\omega_jm_j(y)
\end{equation*}
where $m_j(y)$ is the $x_j$-marginal of $m(\x)$ evaluated at $x_j=y$.
The prior forecast $p(y)$ is a linear pool of $h_0(y)$ and the {\em marginal expectations} of each of the model forecast densities.

Simple extensions allow $\mD$ to inject adjustments for expected biases or aspects of mis-calibration in each of the model predictions. The simplest, for example, is to modify the synthesis function to 
\begin{equation*}
    \alpha(y|\x) = \omega_0h_0(y) + \sum_{\jj}\omega_j\djb(y),
\end{equation*}
where $\mD$ specifies expected bias terms $\beta_j$. This results in the synthesis function
\begin{equation*}
    p(y|\mH) = \omega_0h_0(y) + \sum_{\jj}\omega_jh_j(y+\beta_j), 
\end{equation*}
i.e.,  $\mD$ has simply  adjusted the locations of model forecast distributions to address  expected biases.

\subsection{Model-Specific Outcome-Dependent Weights}
\subsubsection{General Framework.} 
\label{sec:framework}

The first generalization to outcome-dependent weights allows the mixture probabilities $\omega_j$ to depend on the latent forecast location $x_j$ via
\begin{equation}
    \label{eq:specific_weights}
    \alpha(y|\x) = \omega_0(\x)h_0(y) + \sum_{j=1:J}\omega_j(x_j)\delta_{x_j}(y)
\end{equation}
where $\omega_0(\x)=1-\sum_{j=1:J}\omega_j(x_j)$. 
Of course, \eqno{constant_weights} is a special  case.
\Eqno{bps} then yields $\mD$'s  synthesis 
\begin{equation*}
    p(y|\mH) = c_0 h_0(y) + \sum_{\jj}c_j h_j'(y)
\end{equation*}
where $c_j=\int\omega_j(y)h_j(y)dy$ for each model $\jj$, $c_0=1-\sum_\jj c_j$, and $h_j'(y)=\omega_j(y)h_j(y)/c_j$. 
In this synthesis,  $\mD$ has adjusted each density $h_j(y)$ to a re-calibrated $h_j'(y)$.
BPS provides an explicit framework for calibration, and importantly, also delineates the assumptions and conditions required to maintain calibration coherence.
Notably, the weights $w_j(\cdot)$ could be functions not only of the $x_j$, but perhaps also of additional relevant covariates, model scores, or decision-related measures, as in recent developments in~\cite{LavineLindonWest2021avs} and~\cite{TallmanWest2022}. 

\subsubsection{Example: Gaussian Weights.}
A first example is inspired by~\cite{Fawcett2014}. Here different models are regarded as more or less informative in different regions of the outcome variable (e.g., bear markets with negative values vs.\ bull markets with positive values). Take
\begin{equation*}
    \omega_j(x_j) = q_j\exp(-(x_j-\mu_j)^2/(2\sigma_j^2))
\end{equation*}
for $\jj$, where the $q_j$ are a set of $J$ {\em base synthesis weights} that sum to 1, and  $\omega_0(\x)=1-\sum_{\jj}\omega_j(x_j).$  Write $\q$ for the vector of base synthesis weights. 
Here, the overall weight on $\mM_j$ takes a maximal value of $q_j$ at $x_j=\mu_j$ and decreases as $x_j$ moves away from $\mu_j$, with the rate of decrease controlled by $\sigma^2_j$.
In this way, $\mD$ encodes that $\mM_j$ is trusted most near $\mu_j$.

\begin{figure}[htbp!]
\subfloat[$h(y)=N(y|1,0.1)$]{%
  \includegraphics[clip,width=0.5\columnwidth]{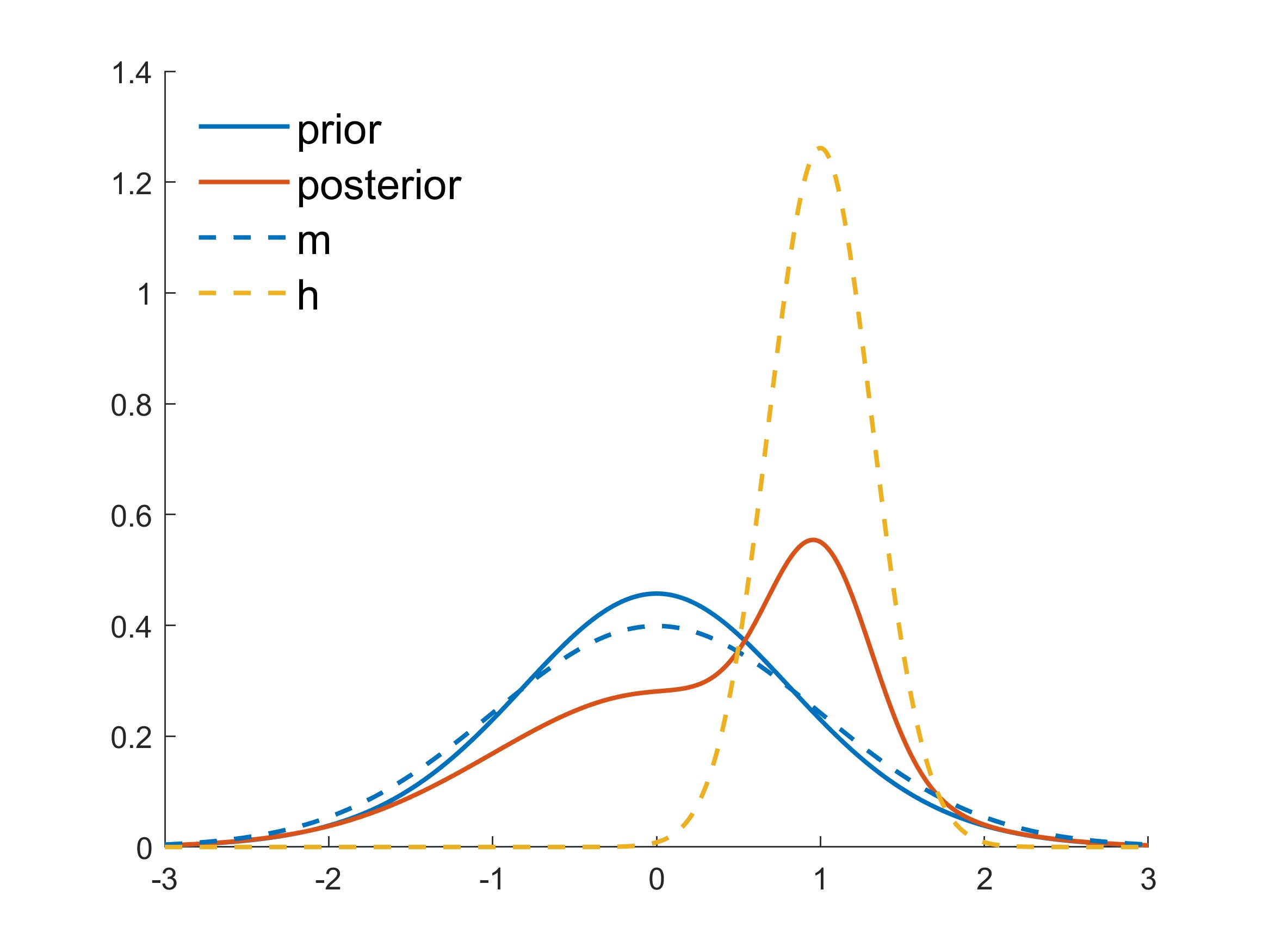}%
  \label{fig:lowf}
}
\subfloat[$h(y)=N(y|2,0.1)$]{%
  \includegraphics[clip,width=0.5\columnwidth]{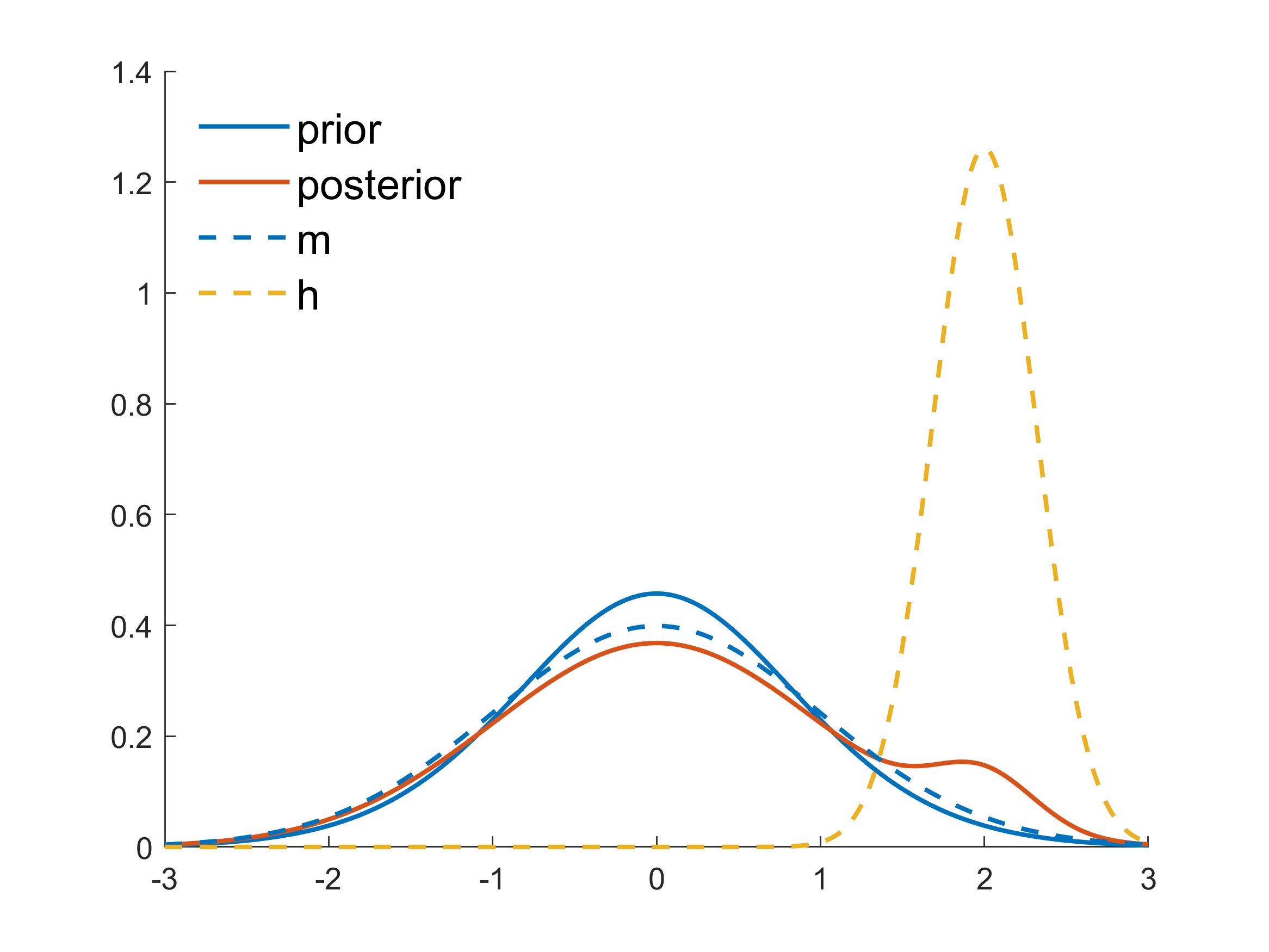}%
  \label{fig:highf}
}
\caption{Posterior updates under two different model forecasts in the single model setting with $q=0.5$. The baseline density $h_0(y)$ and the expected density $m(y)$ are standard normal, which under outcome-dependent mixing weights produces a slightly more diffuse prior according to \eqno{consistency}. Figure~\ref{fig:lowf} displays the results when the model prediction is $N(1,0.1)$, and Figure~\ref{fig:highf} when it is $N(2,0.1)$.
In each case, the model forecast is relatively precise. When $f=1$ (near $\mu=0$) $\mD$  takes this as a good sign and places higher weight on the model prediction.  However, when $f=2$, the model forecast is viewed as \lq\lq less expected'' and is given lower weight.}
\label{fig:updateJ1}
\end{figure}

Consider an example with a single model,  $J=1$, and drop the then-superfluous $j$ subscript for clarity. 
Set $h_0(y)=N(y|0,1)$ and $m(x)=N(x|\mu,\sigma^2)$ so that $\mD$
expects the model to generate forecasts that, relative to baseline, are location-shifted by a factor $\mu$ and scale-adjusted through $\sigma$. 
Then $\omega(x)=q\exp\{-(x-\mu)^2/(2\sigma^2)\}$.
Suppose the model now presents $h(y) = N(y|f,s)$ for some point forecast $f$ and variance $s$.
The effect of $\omega(y)$ on $h(y)$ is to down-weight the portions of $h(y)$ that are further from $\mu$, resulting in the reweighted model density 
\begin{equation*}
h'(y) = N(y|a_1\mu+a_2f,a_1a_2(\sigma^2+s))
\end{equation*}
where $a_1=s/(\sigma^2+s)$ and $a_2=\sigma^2/(\sigma^2+s)$.
So $h'(y)$ is a compromise between the forecast that $\mD$ expects and the forecast that $\mM$ provides.
The weight on $h'(y)$ in the mixture form is
\begin{equation*}
 q\sqrt{\sigma^2/(\sigma^2+s)}\exp\{-(f-\mu)^2/(2(\sigma^2+s))\},
\end{equation*}
further emphasizing how lower weight is given to the adjusted model density as $f$ moves away from $\mu$.
Figure~\ref{fig:updateJ1} demonstrates $\mD$'s prior-to-posterior update given two different forecasts $h(\cdot).$

\subsubsection{Gaussian Well Weights.}
\label{sec:gaussian_well}
In a different applied setting, $\mD$ might be concerned about the reliability of the model forecasts from one model $\mM_j,$ and aim to down-weight that model relative to others and the baseline in regions relatively favored by $\mM_j.$   A synthesis function that reflects this is the Gaussian well form
\begin{equation*}
    \omega_j(x_j) \propto 1 - \exp(-(x_j-\mu_j)^2/(2\sigma_j^2)).
\end{equation*}
This example emphasizes emphasizes the flexibility of of choices in the mixture-based BPS framework.
Details are not explored here, but weighting with Gaussian wells is further developed in section~\ref{sec:herding}.
 
\subsection{Cross-Model Weights}

\subsubsection{General Framework: Model Dependencies.} 

Generalizations have model weight functions that depend on the  full vector $\x$ rather than just the individual $x_j$ in weighting $\mM_j.$ 
The general form is
\begin{equation}
    \label{eq:cross_weights}
    \alpha(y|\x) = \omega_0(\x)h_0(y)+\sum_{\jj}\omega_j(\x)\delta_{x_j}(y)
\end{equation}
where $\omega_{0:J}(\x)$ are non-negative and sum to one for each $\x$.
\Eqnos{constant_weights}{specific_weights} are special cases.
The resulting predictive synthesis is
\begin{equation*}
    \label{eq:wj}
 p(y|\mH) =  c_0h_0(y) + \sum_\jj c_jh_j'(y)
\end{equation*}
where each $h_j'(y) =  w_j(y)h_j(y)/c_j $ with 
\begin{equation}
    \label{eq:wj}
    w_j(y) = \int \tilde \omega_j(\xj, y)\prod_{i\ne j}h_i(x_i)dx_i
\end{equation}
in which  $\tilde\omega_j(\xj, y)$ is $\omega_j(\x)$ evaluated at $x_j=y$, and the $c_j=\int_y w_j(y)h_j(y)dy=\int_\x\omega_j(\x)h(\x)d\x$ are normalizing constants. 

Allowing $\omega_j(\cdot)$ to depend on the full vector of latent states $\x$ rather than just $x_j$  generalizes and 
extends the interpretation of outcome-dependent weight pooling. 
\Eqno{wj} defines opportunity to weight $\mM_j$ predictions  given the set of forecasts from other models, as well as in terms of its own specific biases and expected prediction accuracy.  This allows adjustments for potential outlier forecasts, and for dependencies-- including expected herding behavior, for example-- among models. For instance, if three models using similar data provide similar forecasts, equal weights may suffice.
If one of the three provides a forecast different from the other two, $\mD$ may wish to give it 50\% of the weight; 
if one model disagrees with 99 others, $\mD$ may choose to ignore it entirely.

\subsubsection{Softmax Weights.}
\label{sec:softmax}
\begin{figure*}[htbp] 
\subfloat[$f_1=-1$, $f_2=1$]{\includegraphics[clip,width=.5\columnwidth]{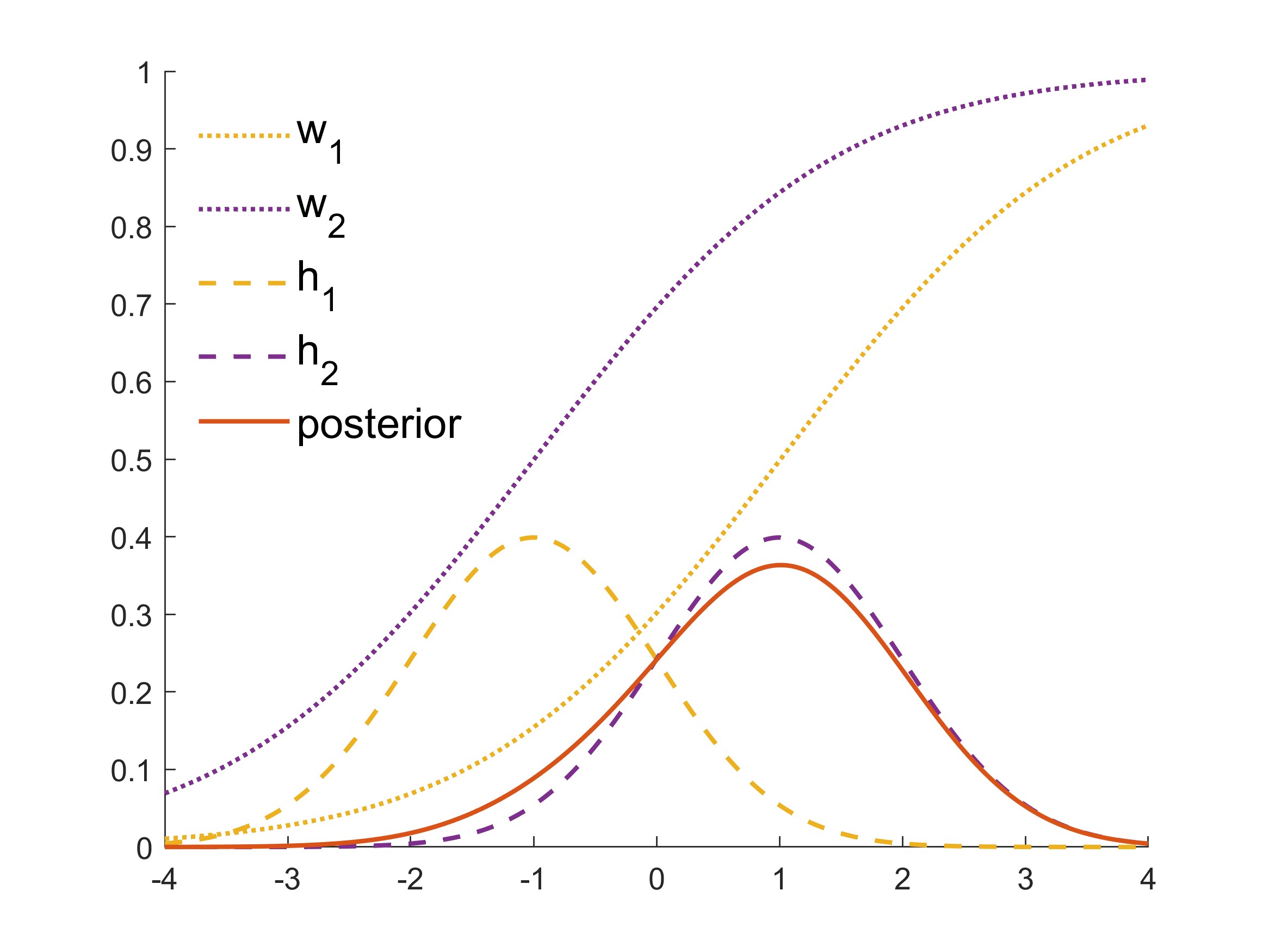}}
\subfloat[$f_1=1$, $f_2=1$]{\includegraphics[clip,width=.5\columnwidth]{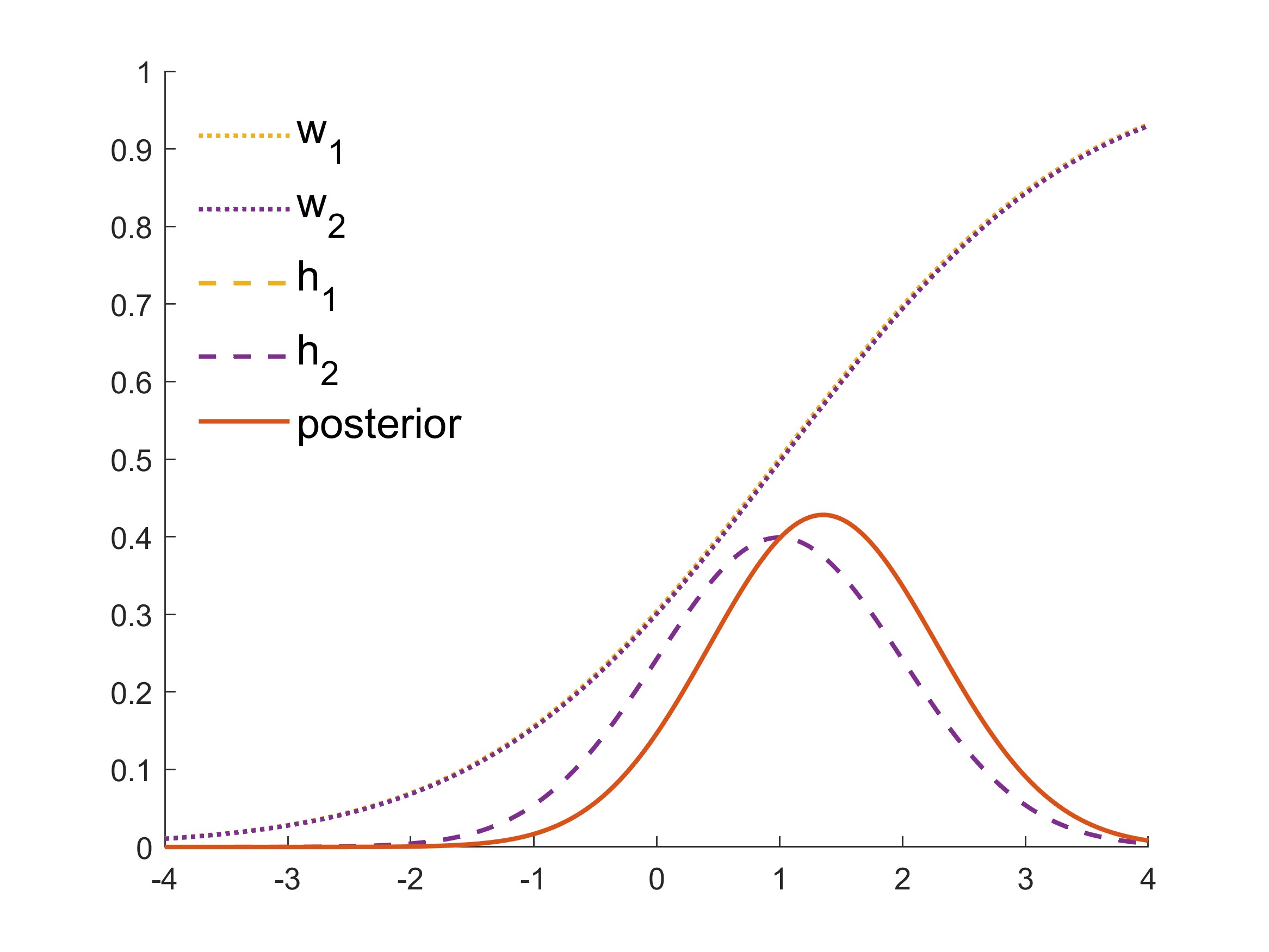}}

\subfloat[$f_1=-1$, $f_2=-1$]{\includegraphics[clip,width=.5\columnwidth]{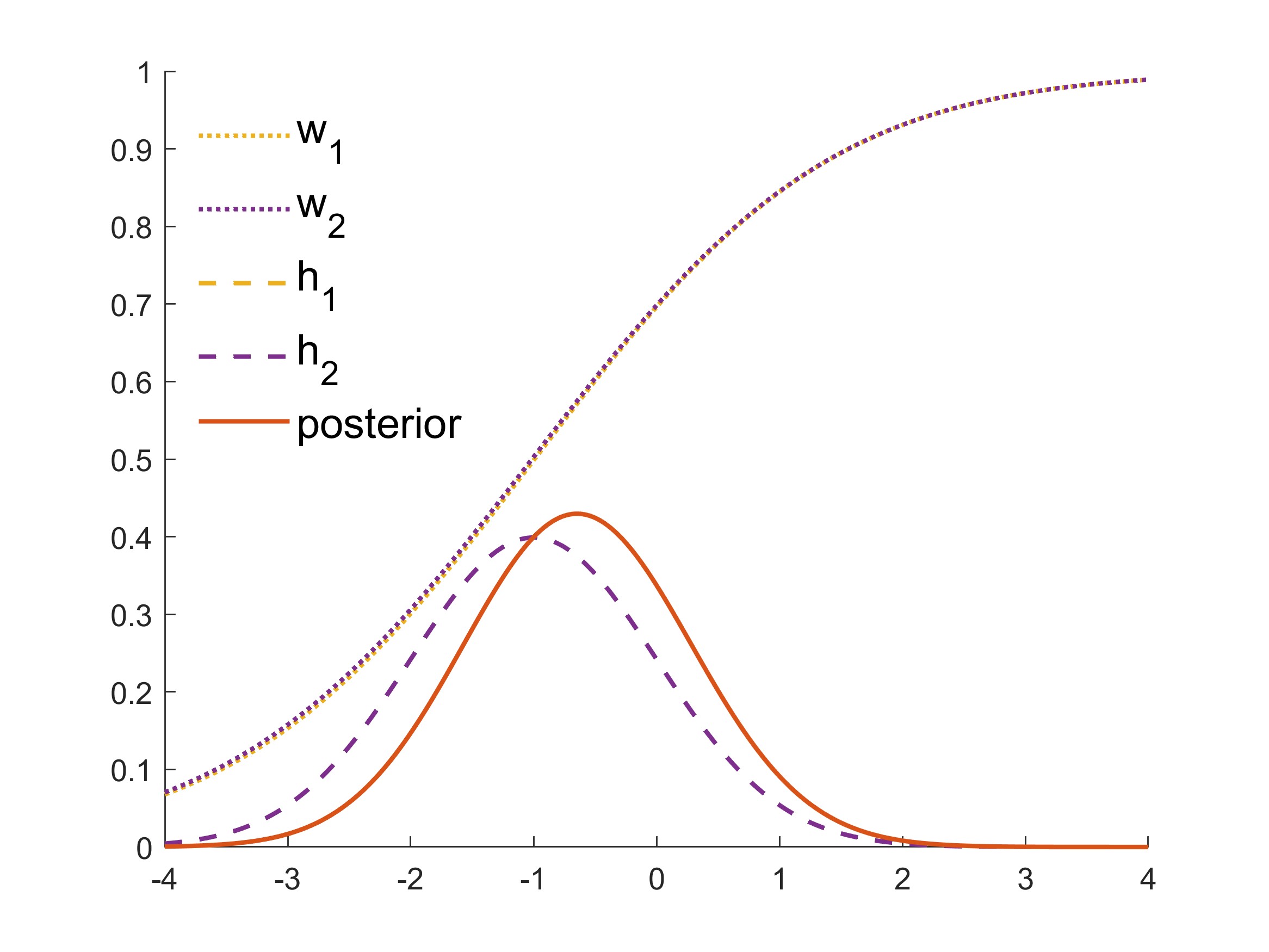}}
\subfloat[$f_1=1$, $f_2=-1$]{\includegraphics[clip,width=.5\columnwidth]{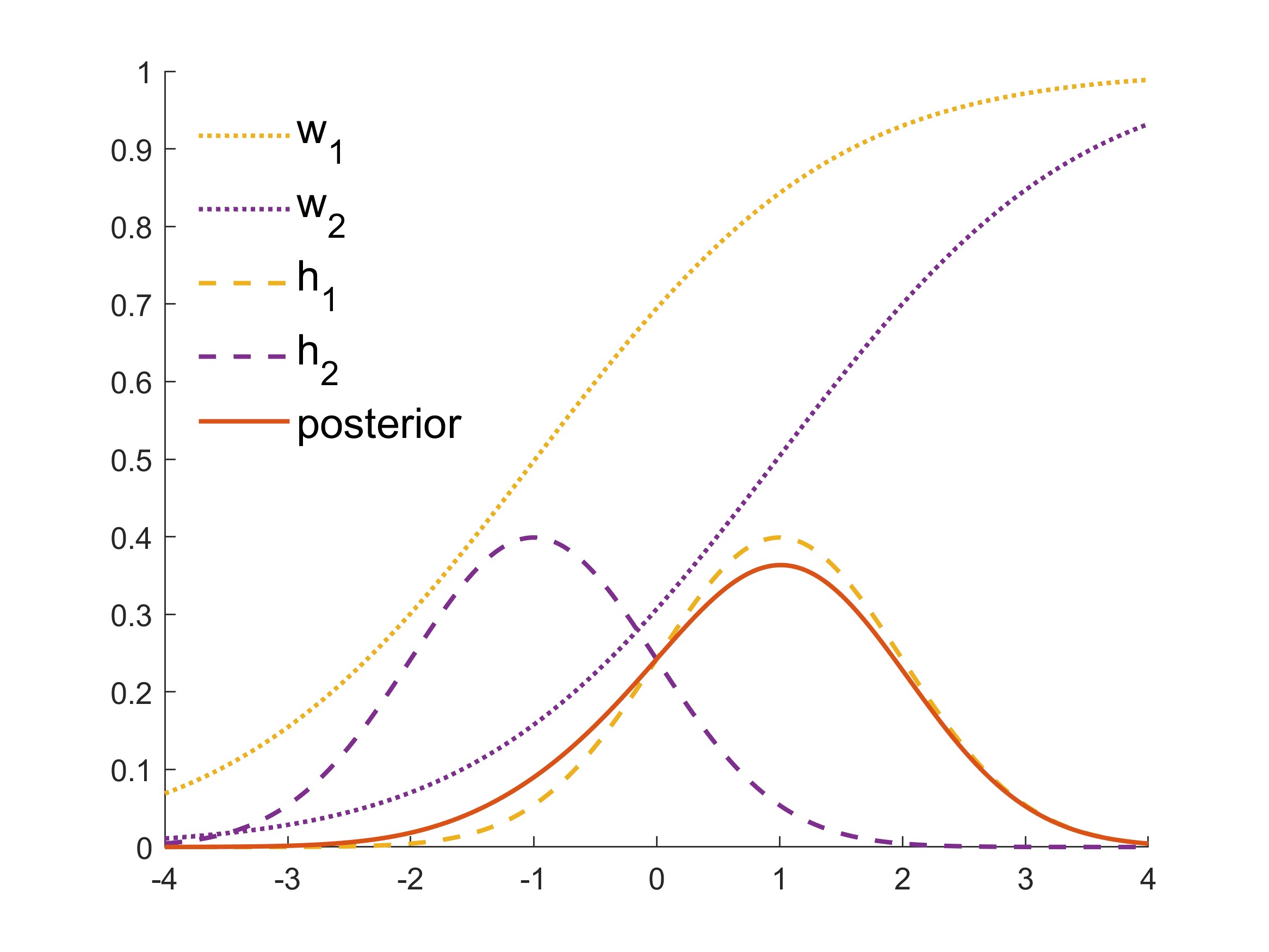}}
\caption{Predictive synthesis in the softmax example of Section~\ref{sec:softmax} with $j=2$ models. Here $\mD$'s predictions favor negative $y$ only when both models favour negative $y$,   When the two model disagree in favoring positive or negative $y$, more weight is given to models supporting larger values of $y$.}
\label{fig:softmax}
\end{figure*}

Consider an example with $J=2$ where
$\omega_1(\x)\propto1$ and $\omega_2(\x)\propto\exp(x_2-x_1)$, with $\omega_0(\x)=0$ so that there is no baseline density.
If $x_1\approx x_2$, $\omega_1(\x)\approx \omega_2(\x)\approx0.5$.
If $x_1>x_2$, $\omega_1(\x)>\omega_2(\x)$, and vice versa.
In this specification, $\mD$ prefers higher forecasts-- higher weight is given to the higher density, regardless of the model that provides it.
Figure~\ref{fig:softmax} shows $p(y|\mH)$ and $w_j(y)$ when $h_j(y)=N(f_j, 1)$ for $f_j=\pm1$.
Note that $\mD$ ignores low forecasts unless both models provide low forecasts.

\subsubsection{Weighting for Consensus.\label{sec:consensus}}
In setting the weight function, the general way to phrase the question is ``How much weight should be placed on $\delta_{x_j}(y)$, given the entire vector $\x$?"
Posed another way, ``For given $\xj$, how much weight should be placed on $\delta_{x_j}(y)$, as a function of $x_j$?" 
Conceptually in terms of values of the latent model states, $\mD$ may wish to ignore a value $x_j$ that disagrees with the others, i.e.\ it falls outside of a ``consensus'' of the remaining $J-1$ models.
One way to do this is to decrease the weight on $\dj(y)$ around $E[x_j|\xj]$, which is defined through the density $m(\x)$. 

A key example introduces a multivariate normal $m(\x)$ for the vector of latent model states, namely
$m(\x) = N(\x|\bmu,\bSigma)$  with mean and covariance matrix $(\bmu,\bSigma).$  This allows for the representation of expected cross-model dependencies through $\bSigma.$   This example is extended to the time-varying setting in Section~\ref{sec:CWTVBPS}, and explored in the subsequent example in Section~\ref{sec:application}. 

Under this choice of $m(\x),$ the implied complete conditional $p(x_j|\x_{-j})$ is normal with $E[x_j|\xj] = \mu_j + \bgamma_j'(\xj-\bmu_{-j})$ where the regression vector $\bgamma_j$ is  implied by $\bSigma; $ write $\nu_j$ for the corresponding conditional variance of $(x_j|\xj)$.  A natural choice for
  $\omega_j(\x)$ involves  the kernel of the implied conditional normal p.d.f., taking
\begin{equation}
    \label{eq:consensus}
    \omega_j(\x) = q_j\exp(-e_j^2/(2\nu_j)), \quad\jj,
\end{equation}
where $e_j = x_j - E[x_j|\x_{-j}],$  the deviation of $x_j$ from its point prediction based on the latent states of the other models,  and
with, as before,  $q_j$  defining base synthesis weights. 
If $m(\x)$ is specified so that the latent states are positively correlated, lower 
weight is given to $\dj(y)$ when $e_j$ is large in absolute value, i.e., when $x_j$ is far from its conditional expectation based on the latent states of the other models.

\subsubsection{Weighting for Herding.\label{sec:herding}}
\label{sec:herding}
\begin{figure}[tb]
\centering
\includegraphics[clip,width=.5\columnwidth]{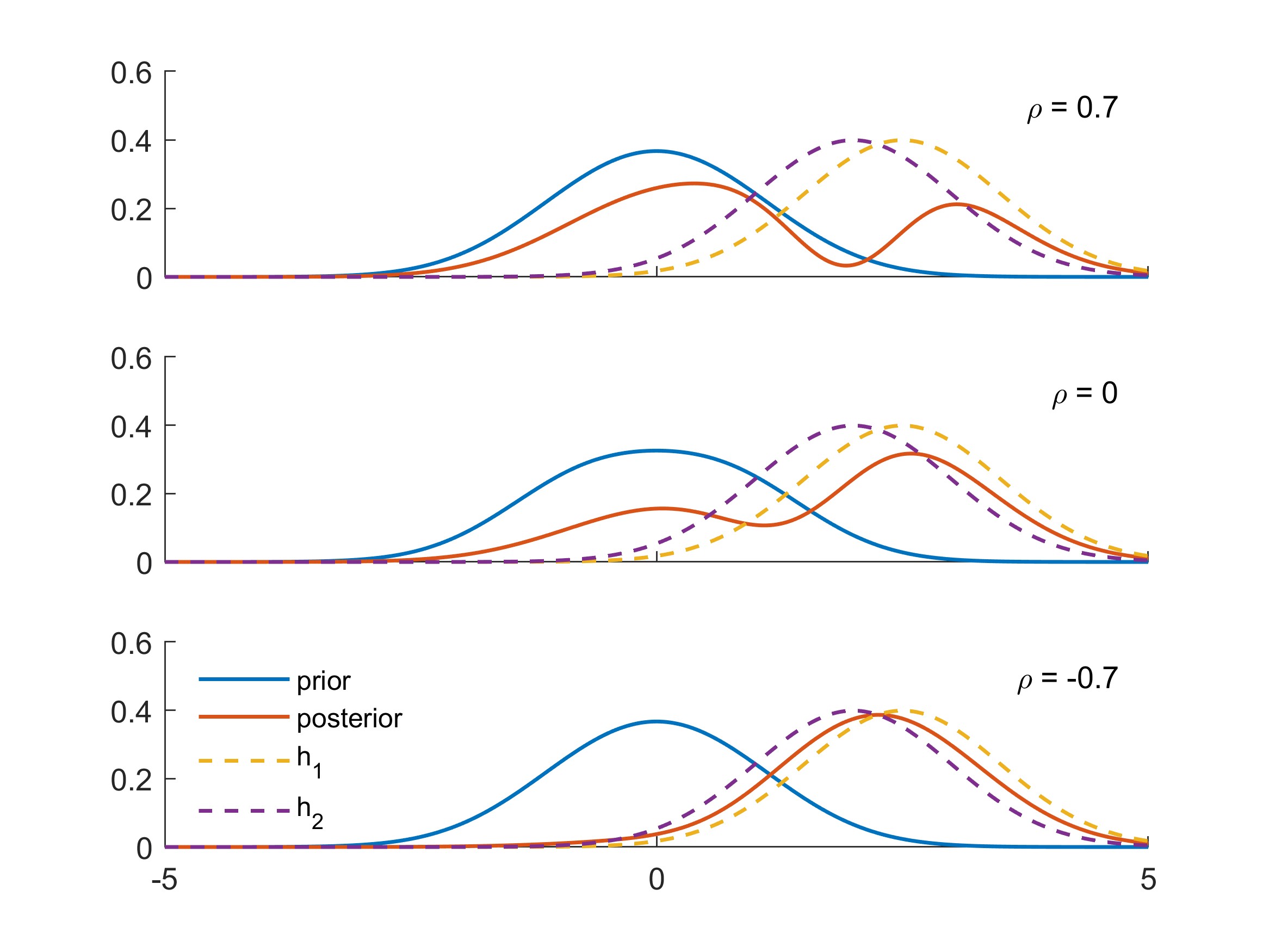}
\caption{Synthesizing similar model densities using weights based on conditional Gaussian wells (Section~\ref{sec:herding}). In each case, the expectation function $m(\x)$ is bivariate normal with standard margins, while the correlation takes values in $\{0.7,0,-0.7\}$. Lower weight is assigned to similar model forecasts when $\mD$ expects them to agree, and the weight increases as the set of forecasts becomes increasingly ``unexpected.''}
\label{fig:well_example}
\end{figure}

Addressing cross-model dependencies and the herding issue (positive dependencies among model predictions) more directly, $\mD$   may wish to decrease weight on a  predictive \lq\lq consensus''. This can be targeted using a number of synthesis function choices, including the example
\begin{equation}\label{eq:herding}
    \omega_j(\x) = q_j(1 - d\exp(-e_j^2/(2\nu_j))), \quad\jj,
\end{equation}
with the $e_j, \nu_j$ as in Section~\ref{sec:consensus}, and where $d$ denotes the depth of the conditional Gaussian well as in Section~\ref{sec:gaussian_well}. As a result, the weight on $\mM_j$ will be increased as $e_j$ increases in absolute value, as $\mD$ seeks a diversity of forecasts rather than a consensus.
Figure~\ref{fig:well_example} illustrates the results of weighting similar model densities using conditional wells under different assumptions about model dependencies.
When models are expected to agree, similar forecasts are weighted lower; hence the analysis naturally discounts positively dependent model forecasts, accounting for herding. In contrast, when they are expected to disagree with negative dependence in the synthesis,   similar observed forecast distributions are more highly weighted;  this is again natural in reflecting agreement in what is expected to be an antithetical setting.

\section{Dynamic Mixture BPS for Time Series\label{sec:TSBPS} }

\subsection{Time Series Context}
\label{sec:timeseries}
Common applications are in time series forecasting, where $\mD$ receives predictive distributions from the same set of models repeatedly over time.  
 Here a parameterized synthesis function may be time-varying. Then $\mD$ sequentially updates information relevant to the synthesis parameters to reflect evolving predictive  accuracy of the models, and perceptions of bias and dependence between models, all of which may vary in time. 

Adding subscript $t$ to denote equally spaced time, focus first on 1-step ahead prediction of a scalar time series. At each time $t-1$, $\mD$ forecasts $y_t$ based on historical information and new predictive densities supplied by the set of models, each predicting $y_t.$ 
 $\mD$'s analysis is implicitly conditional on the time $t-1$ filtration consisting of the observed model forecasts $\mH_{1:t-1}$ and data $y_{1:t-1}$, though for notational clarity this is not made explicit here.   

\subsection{A Dynamic Combined Weighting BPS Model\label{sec:CWTVBPS}} 
  
The synthesis function of~\eqno{cross_weights} is  generalized to reflect time dependence throughout: the 
synthesis weights are now $\omega_{jt}(\cdot) = q_{jt} \alpha_{jt}(\cdot)$ with $\alpha_{jt}(\cdot)\in[0,1]$, and the model is 
extended to including potentially time-varying model biases $\beta_{jt}$. This results in 
\begin{equation}
    \label{eq:cross_weights_time}
    \alpha(y_t|\x_t) = \omega_{0t}(\x_t)h_{0t}(y_t)+\sum_{\jj}\omega_{jt}(\x_t)\delta_{x_{jt}-\beta_{jt}}(y_t)
\end{equation}
with implicit latent factors $\x_t$ that are now also time-dependent, i.e., define a vector time series of dynamic latent factors.  

Structuring uses time-dependent extension of~\eqno{consensus}. 
The time-specific $m_t(\x_t) = N(\bmu_t,\bSigma_t)$ have univariate complete conditionals 
that  imply $w_{jt}(\cdot)$.    Cross-model dependencies and their evolution in time are reflected in the $\bSigma_t$.  Coupled with this, 
evolving model-specific biases are reflected in the specification 
 $\bmu_t=f_{0t}+\bbeta_t$, where $f_{0t}$ is the known mean of the specified baseline density $h_{0t}(y_t)$, 
and $\bbeta_t$ is a $J-$vector of bias terms $\beta_{jt}$. Thus 
 $\mD$ expects model $\mM_j$ to have a bias of $\beta_{jt}$ relative to $f_{0t}$ in forecasting $y_t,$ with
time-variation accommodated.    The bias term acts directly to adjust the mixture locations in the synthesis function, as in the static example at the end of Section~\ref{sec:constantmixBPS}.   This translates parameter learning from $(\bmu_t,\bSigma_t)$ to $(\bbeta_t,\bSigma_t)$, along with the 
vector $\q_t$ of time $t$ base synthesis weights $q_{jt}$.

The analysis to follow assumes that, at time $t-1$,  accrued historical information leads $\mD$ to summarize the time $t-1$ posterior for model parameters as follows:   $(\bbeta_{t-1},\bSigma_{t-1})$ have a normal, inverse-Wishart (NIW) distribution independently of $\q_{t-1}$, while the latter has a  Dirichlet distribution.
Time variation in parameters is defined using standard discount factor methods for Bayesian dynamic modeling.  In moving to time $t$, the parameters
 $(\bbeta_{t-1},\bSigma_{t-1})$ evolve to $(\bbeta_t,\bSigma_t)$ and the implied time $t$ prior-- before observing $y_t$-- is also NIW but with increased uncertainty representing potential changes through the evolution. Standard discount theory for dynamic linear modeling underlies 
 this~(\citealp{West1997},  chapter 16;  \citealp{PradoFerreiraWest2021}, chapter 10).
 In parallel and independently, $\q_t$ evolves according to a dynamic Dirichlet model: $\q_{t-1}$ evolves to $\q_t$ and the implied prior-- before observing $y_t$--  is also Dirichlet but with a precision parameter that is reduced by a discount factor to represent increased uncertainty. 

\subsection{Sequential Model Analysis and Computation}  
\label{sec:computation}

\subsubsection{Sequential forecasting, filtering and evolution.}
Sequential model analysis involves, at each time $t$, the three steps of forecasting, filtering and subsequent evolution to $t+1.$ First, at time $t-1$, forecast or predict $y_t$; second, on observing $y_t$ update the prior to posterior over model parameters $\{ \bbeta_t, \bSigma_t; \q_t \}$; third, evolve this time $t$ posterior to the time $t+1$ prior for $\{ \bbeta_{t+1}, \bSigma_{t+1}; \q_{t+1} \}$. The process then repeats over future time periods.   

As noted above, the time $t$ prior distributions are assumed by $\mD$ as independent 
NIW and Dirichlet. The BPS model defines $p(y_t|\bbeta_t,\bSigma_t,\q_t)$ implicitly through the mixture over~\eqno{cross_weights_time} with respect to the model set input product $\prod_{\jj} h_{jt}(x_{jt}).$ Whatever the $h_{jt}(\cdot)$ may be,  the complexity of analytic form of the BPS weights $w_{jt}(\cdot)$ generally obviates any analytic evaluation of predictive and posterior/filtered quantities of interest. Hence much of the analysis is simulation-based for both prediction and posterior analysis.  Then, simulation samples from the time $t$ posterior define the basis for constraints to evolve to the constrained NIW and Dirichlet priors at time $t+1$.  The three steps are summarized as follows. 

\subsubsection{One-step prediction at time $t-1.$}
This is trivial via direct Monte Carlo: (i) simulate parameters from the NIW 
prior for $\{\bbeta_t,\bSigma_t\}$ and $\q_t$ from its Dirichlet prior; (ii) simulate independent draws of the model latent states $x_{jt} \sim h_{jt}(\cdot)$;  conditional on these synthetic values, simulate $y_t$ from~\eqno{cross_weights_time}.   Repeat to generate a Monte Carlo random sample from the one-step ahead forecast distribution;  summarize as desired.


\subsubsection{Prior-to-posterior update at time $t$.} 

At each time $t$ on observing outcome $y_t$,  a structured Gibbs sampling-style 
Monte Carlo Markov chain (MCMC) sampler defines a simulation approach to evaluation of the 
posterior for model parameters $\{ \bbeta_t, \bSigma_t; \q_t \}$ and the latent model states $\x_t$ jointly.   
Within each overall MCMC iteration,  components of this sampler involve exact simulation from relevant conditional posteriors that are analytically tractable, 
while other components exploit accept/reject sampling. Further details are in Appendix~\ref{sec:suppGibbs}.

To complete the update step, $\mD$ uses the Monte Carlo posterior sample to define the analytic posterior NIW and Dirichlet distributions required for evolution to the next time point $t+1.$   This is done via variational Bayes as  in~\cite{GruberWest2016BA,GruberWest2017ECOSTA} in related contexts. Specifically,  the parameters of a posterior NIW and Dirichlet for $\{\bbeta_t,\bSigma_t; \q_t\}$ given $y_t$ are computed by minimizing the  K\"{u}llback-Leibler divergence of  the resulting, analytic form from the empirical posterior represented by its Monte Carlo sample. Further details are summarized in Appendix~\ref{sec:vb}.
 
\subsubsection{Evolution from $t$ to $t+1$.} 

 In moving to time $t+1$, the NIW and Dirichlet posteriors are modified with discount factors to define the implied time $t+1$ prior for 
 $(\bbeta_{t+1},\bSigma_{t+1}; \q_{t+1})$; the discount evolution simply increases uncertainty in the distributions in moving ahead one time point, precisely as already discussed above for the time $t-1$ to $t$ evolution. See Appendix~\ref{sec:Appdiscounts}. 
 
\section{Time Series Example}
\label{sec:application}

\subsection{FX Time Series Setting} 

\begin{figure}[t!]
\centering
\includegraphics[clip,width=.5\columnwidth]{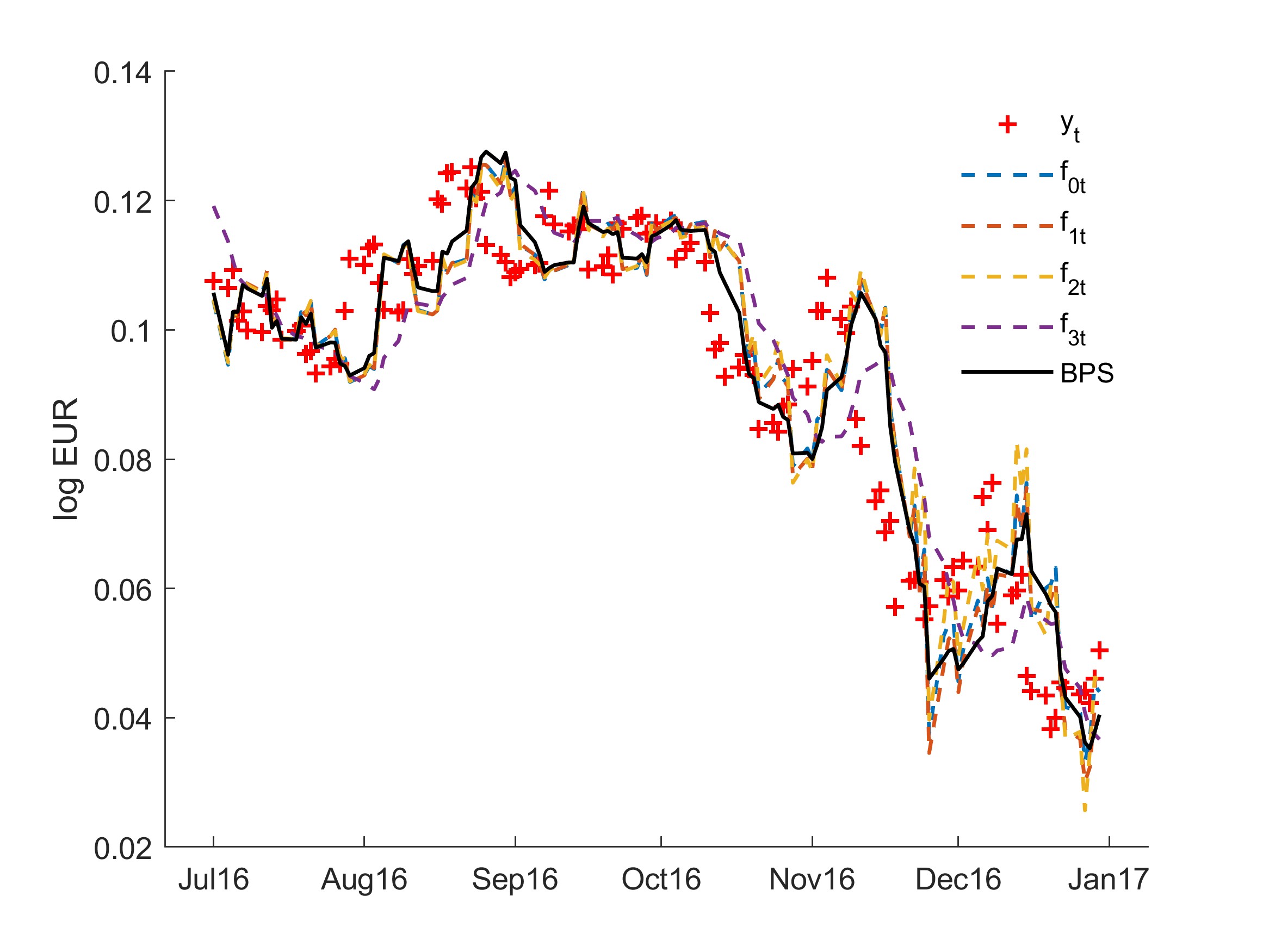}
\caption{Euro log-price daily data, 5-step ahead model point forecasts and
BPS point forecasts.}
\label{fig:data}
\end{figure}

The BPS model and analysis of Section~\ref{sec:TSBPS} is explored in a study of a daily FX (foreign exchange) currency prices, namely that of the Euro relative to the US\$ over a period of six months.   Here $y_t$ is the log \$price of the Euro each day over the last six months of 2016, 7/1/2016-12/30/2016, for a total of 130 trading days. This time period includes the U.S. presidential election, which caused some quick FX movements  in mid-November.  The data appear in Figure~\ref{fig:data}. 

In this setting, $\mD$ is interested in predicting 5 days (one working week) ahead. At time $t-5$, after observing $y_{t-5}$, each model generates 
5-step ahead forecast distributions for $y_t$ that are then the inputs to the BPS analysis.  That is, $\mD$ synthesizes forecasts of the specific outcome of interest. This is then repeated each day over the time period of interest.  
With daily FX series, 1-step models are heavily driven by noise and the set of pure time series models here will tend to generate similar 1-day ahead forecasts. Multi-step ahead forecasting allows for more differentiation of model predictive accuracy, and is also much more relevant to financial applications and portfolio decisions~\citep[e.g.][]{ZhaoXieWest2016ASMBI,IrieWest2018portfoliosBA}. Technically, the BPS sequential analysis and computational approach are precisely outlined in Section~\ref{sec:computation}; the difference is simply that of interpretation:  at time $t$, the observation $y_t$  is the outcome that was forecast at time $t-5$, and $(\bbeta_t,\bSigma_t)$ reflect $\mD$'s time $t$ posterior for model biases and dependencies of the 5-day ahead forecasts distributions from the set of models. 

This example serves to illustrate and highlight key aspects of BPS including: (i)  the ability of BPS to identify and adapt to model-specific biases and their changes over time; (ii) to quantify the nature of cross-model dependencies, again with changes over time; (iii) to highlight and adapt to the issue of model set incompleteness; and (iv) to define improved predictions relative to BMA as well as each of the individual models.

\subsection{Model Set and BPS Specification}

Mixture BPS explores and synthesizes a set of $J=3$ dynamic linear models (DLMs) and uses another DLM as the default baseline: 
$\mM_0$ is a time-varying autoregression of order 1, or TVAR(1); 
$\mM_1$ is  TVAR(2); 
$\mM_2$ is TVAR(5); 
$\mM_3$ is a linear growth DLM representing adaptive, locally linear progression over time.
These are standard univariate DLMs and widely used in short-term forecasting and other areas~\cite[][chapters 4 and 5]{PradoFerreiraWest2021}.
Models $\mM_1$ and $\mM_2$ have more predictive potential more than one day ahead as they involve more lagged values of $y_t$ as predictors; 
model $\mM_3$ extrapolates linearly so has similar potential but only on a few days ahead. FX data often shows 2--3 day momentum effects that these models can pick up. In contrast, $\mM_0$ is simply a default that in practice will always score as well as more elaborate models in 1-day ahead forecasting over many time periods, reflecting the fact that short-term daily FX forecasting (with purely time series models) is inherently  very challenging. 

Each day, the analysis of the previous section applies: each model provides predictive distributions 
for the closing (log) price 5 days ahead, and these are dynamically synthesized using the mixture BPS formulation.
Figure~\ref{fig:data} shows resulting point forecasts (5-day ahead forecast means) from each of the models and from BPS.
Importantly, BPS synthesizes and learns from multi-step ahead forecasts, as opposed to extrapolating a combination based on single-step performance. This contrasts with other forecast pooling approaches--  including BMA--  that inherently score models based on 1-step ahead forecasts.

BPS involves choices of: (i) the time-varying  mixture form of~\eqno{cross_weights_time}; (ii) the time $t=0$ initial NIW prior for $(\bbeta_1,\bSigma_1)$ and Dirichlet prior for initial base weights $\q_1$; and (iii) the choice of discount factors that influence how much $\bbeta_t$, $\bSigma_t$, and $\q_t$ vary over time. 
Appendix~\ref{sec:app_details} gives details on all of the above.

\subsection{Aspects of BPS Analysis \label{sec:BPSFXdiscuss}}

\begin{figure}[b!]
\centering
\subfloat[Cross-model correlations.]{%
  \includegraphics[clip,width=.5\columnwidth]{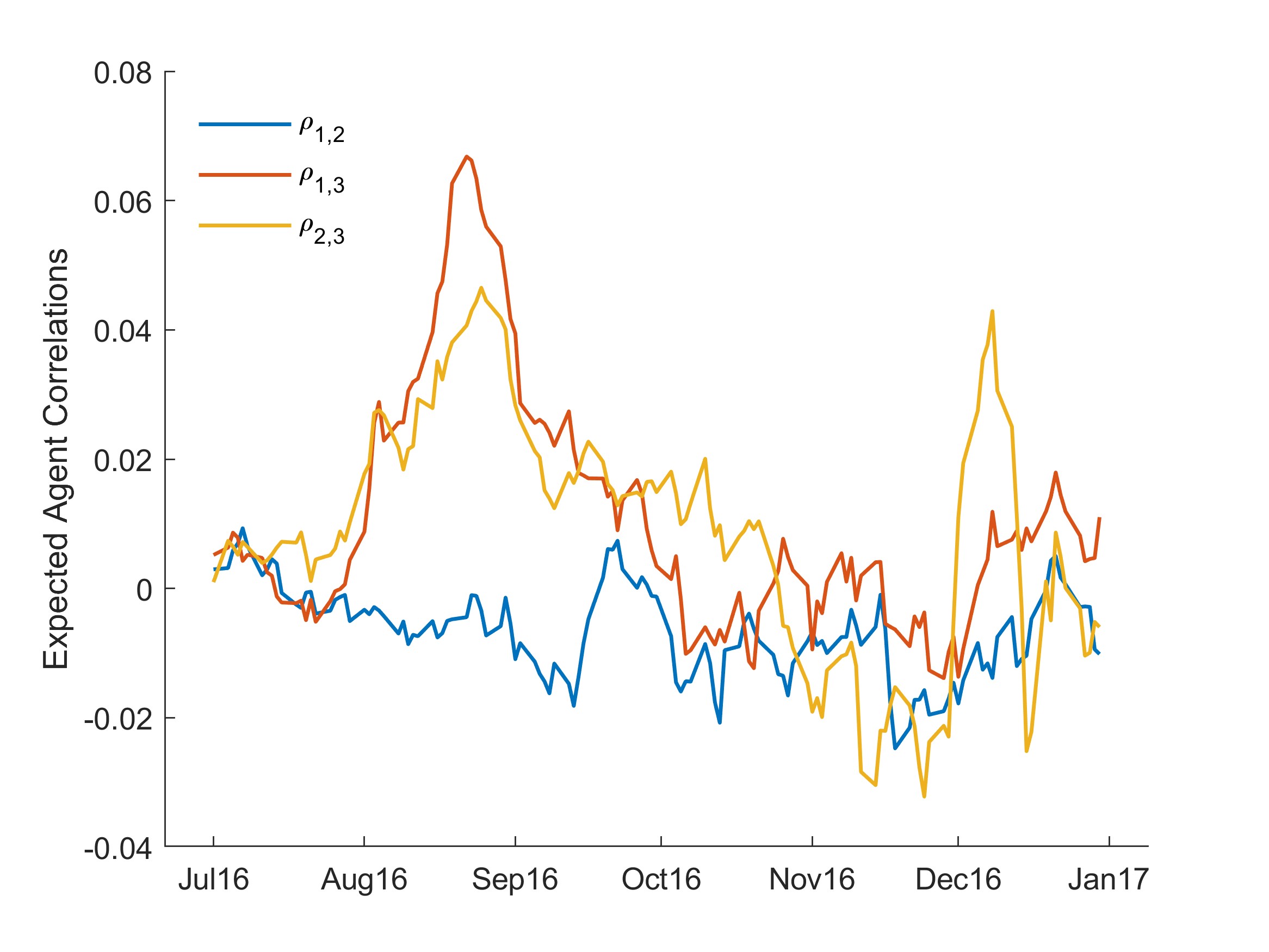}%
  \label{fig:Scorr}
}
\subfloat[Point forecast errors and biases.]{
  \includegraphics[clip,width=.5\columnwidth]{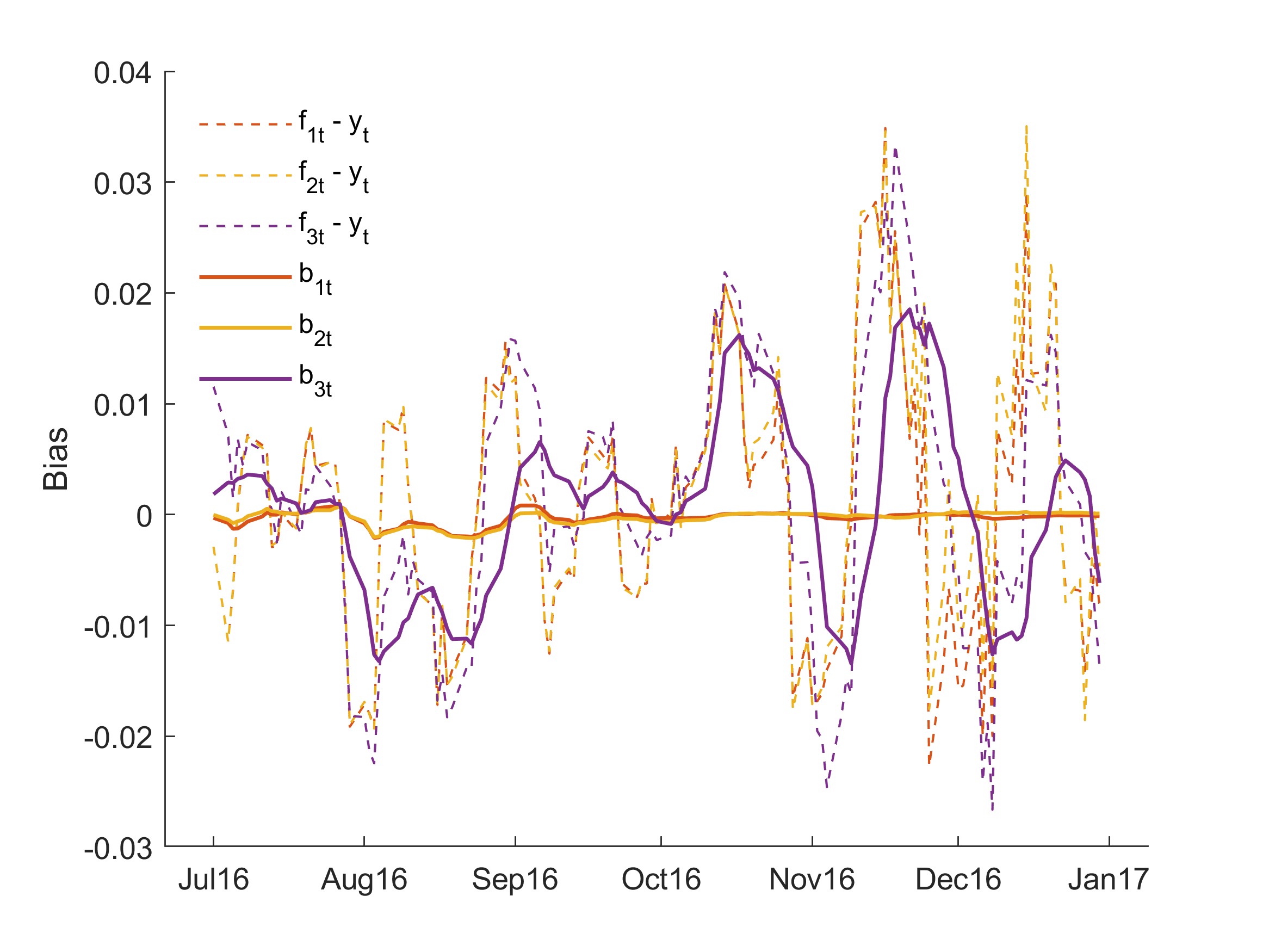}
  \label{fig:bias}
}
\caption{Time trajectories of: (a) filtered posterior means of cross-model correlations underlying $\bSigma_t$ in the BPS synthesis; and (b)  model forecast deviations $f_{jt}-y_t$, and filtered posterior means of the BPS model-specific location biases $\beta_{jt}$, in each of the models $j=\seq 1J.$ }
\end{figure}

As noted above, some main interests are in learning about bias and dependence among models, and in changes over time in these features that BPS is able to represent. The models in the specific model set are expected to perform fairly similarly for most time periods, with time-varying biases, correlations, and scales. Repeat experience with the set of models over time then also builds up a profile of cross-model dependencies. Figure~\ref{fig:Scorr} displays point estimates of correlations represented in the BPS covariance matrix $\bSigma_t$; these are the time $t$ filtered posterior means using the Monte Carlo sample of the posterior for $\bSigma_t$ on day $t$. There is some notable learning over time,  with slightly positive correlations between $\mM_3$ (the locally linear DLM) and each of the TVAR models.
A positive correlation between $\mM_3$ and TVAR models indicates that the model weights are down-weighted when they disagree.
In contrast, BPS learns essentially zero correlation between  $\mM_1$ (TVAR(2)) and $\mM_2$ (TVAR(5)).
As a result, the TVAR models are \emph{not} down-weighted when they disagree; BPS does not require a ``consensus" among TVAR models to assign to them appreciable weights.
All correlations break down around the period of the US election, with some recovery of the slight positive dependency of $\mM_2$ and $\mM_3$ (the longer history models) later in that year. 
The pre- and post-election period was a period of increased uncertainty and consequent volatility in the FX markets, and models with different lag structure respond slightly differently over that period as evidenced by the drop on cross-model correlations. Throughout, interpretable cross-model dependencies are inferred and the time trajectories show the adaptability of the BPS analysis in more volatile time periods.  

Now consider the the vector of bias parameters $\bbeta_t$.  The chosen models  are inherently adaptable to changes over time; this is a main feature of DLMs in terms of addressing model biases. However, the models have discount factor parameters that define their degrees of adaptability. The dynamic BPS model overlays this to allow for systematic, possibly time-varying additional biases in location of prediction distributions through  $\bbeta_t$. The sequential  analysis, illustrated in Figure~\ref{fig:bias}, gives insight into how the BPS model sees biases. This shows time trajectories of posterior means  of the elements of $\bbeta_t$, with evidence of the need for some bias corrections as well as relationships among inferred biases across models over time. The overall BPS analysis integrates inferences on the biases in defining the synthesized predictions at each time, and then in adapting to new, incoming data.

\begin{figure}[t!]
\centering
\subfloat[Dirichlet means {$E[q_{jt}]$}, $j=1:3$.]{%
  \includegraphics[clip,width=.5\columnwidth]{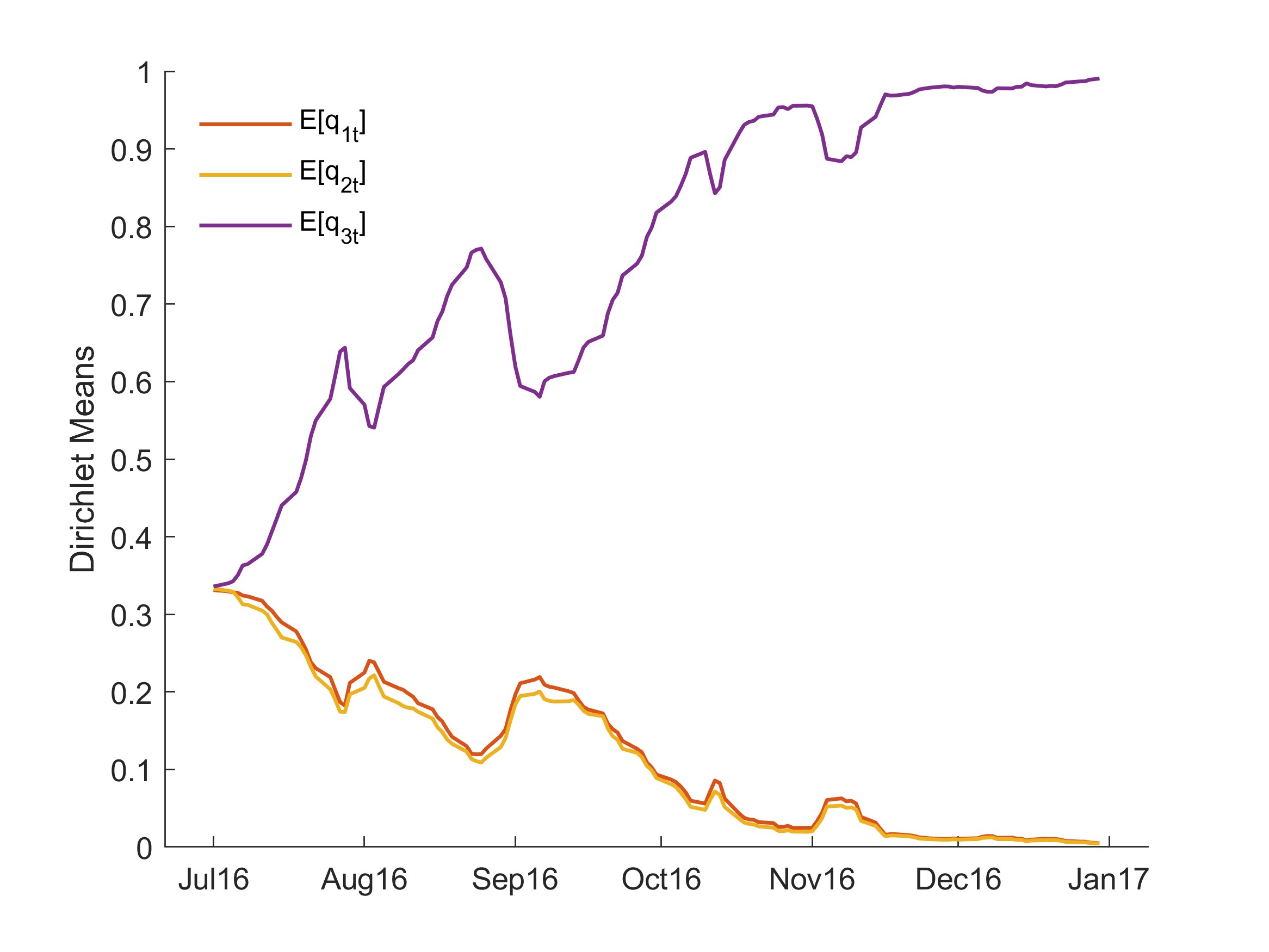}%
  \label{fig:Eq}
}
\subfloat[MCMC model sampling frequencies,  $j=0:3$.]{
  \includegraphics[clip,width=.5\columnwidth]{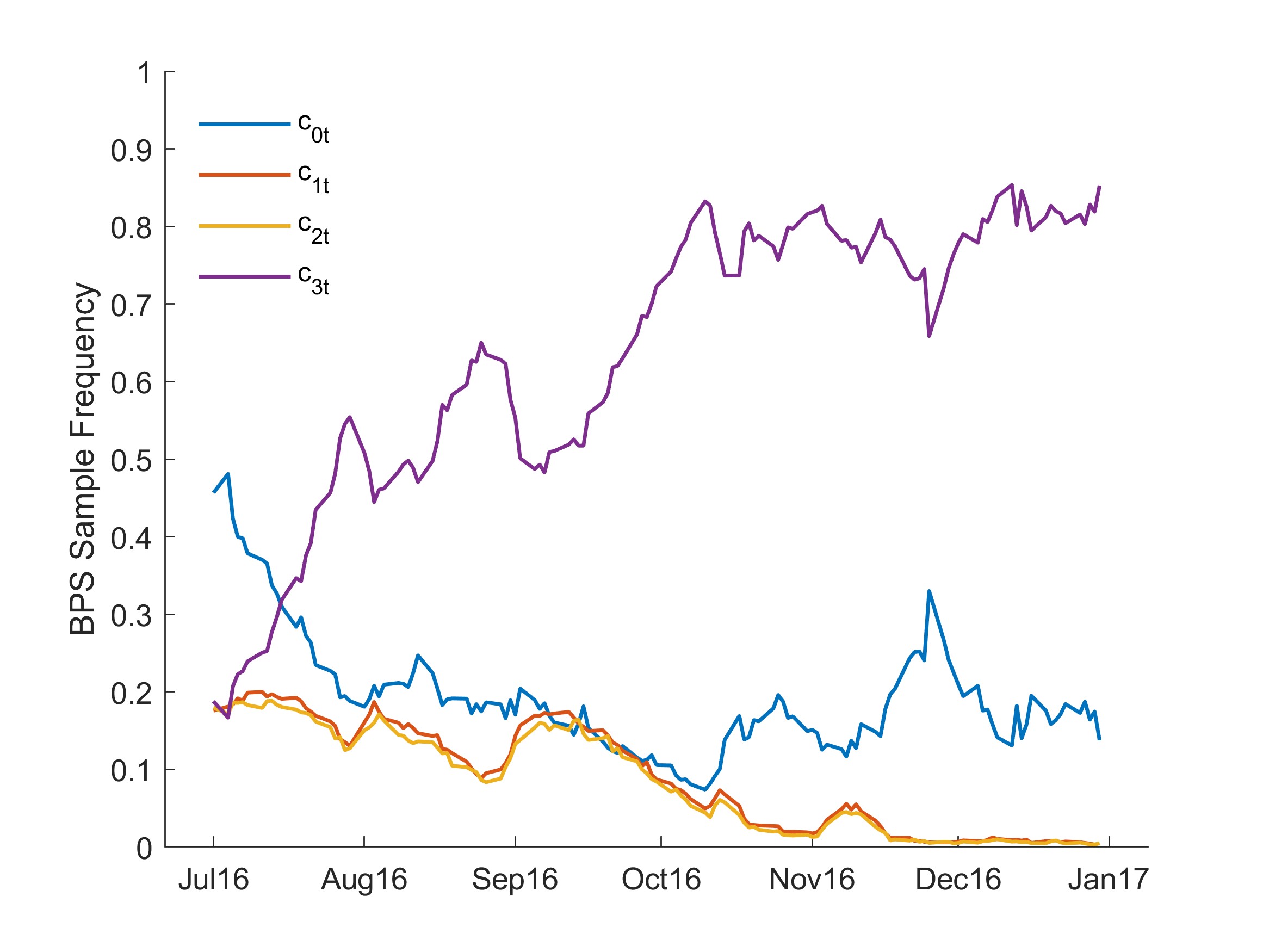}
  \label{fig:zfreq}
}
\caption{Trajectories of BPS model weights: (a) time evolution of filtered posterior means of the base weights $q_{jt}; $  and (b) effective pooling model weights after adjusting for consensus, given by the frequencies of sampling of each of the models in the MCMC analysis.}
\end{figure}

Further evaluation focuses on the BPS baseline weights $\q_t$  and resulting sampling frequencies of the models in the MCMC analysis at each day in the sequential analysis over the full time period.  The filtered trajectories of the sequentially updated posterior Dirichlet distributions for the $\q_t$ is shown in Figure~\ref{fig:Eq}.
Compare these summaries for the baseline weights to the resulting MCMC sampling frequencies of each model in Figure~\ref{fig:zfreq}. The difference between these two figures results wholly from the effect of the BPS outcome-dependent weighting.

\begin{figure}[t!]
\centering
\subfloat[BMA density pooling weights.]{%
  \includegraphics[clip,width=.5\columnwidth]{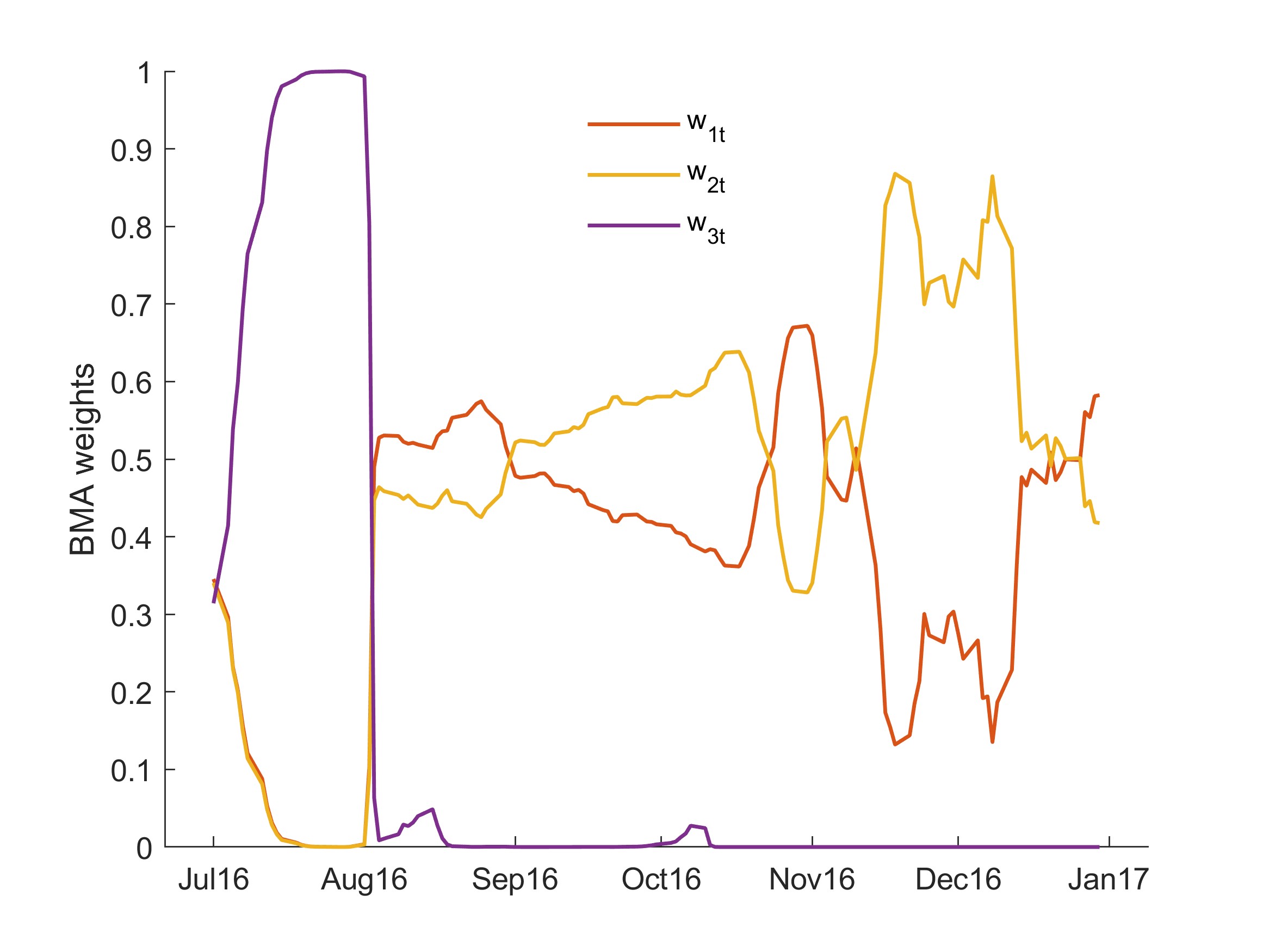}%
  \label{fig:bma}
}
\subfloat[BMAx density pooling weights.]{%
  \includegraphics[clip,width=.5\columnwidth]{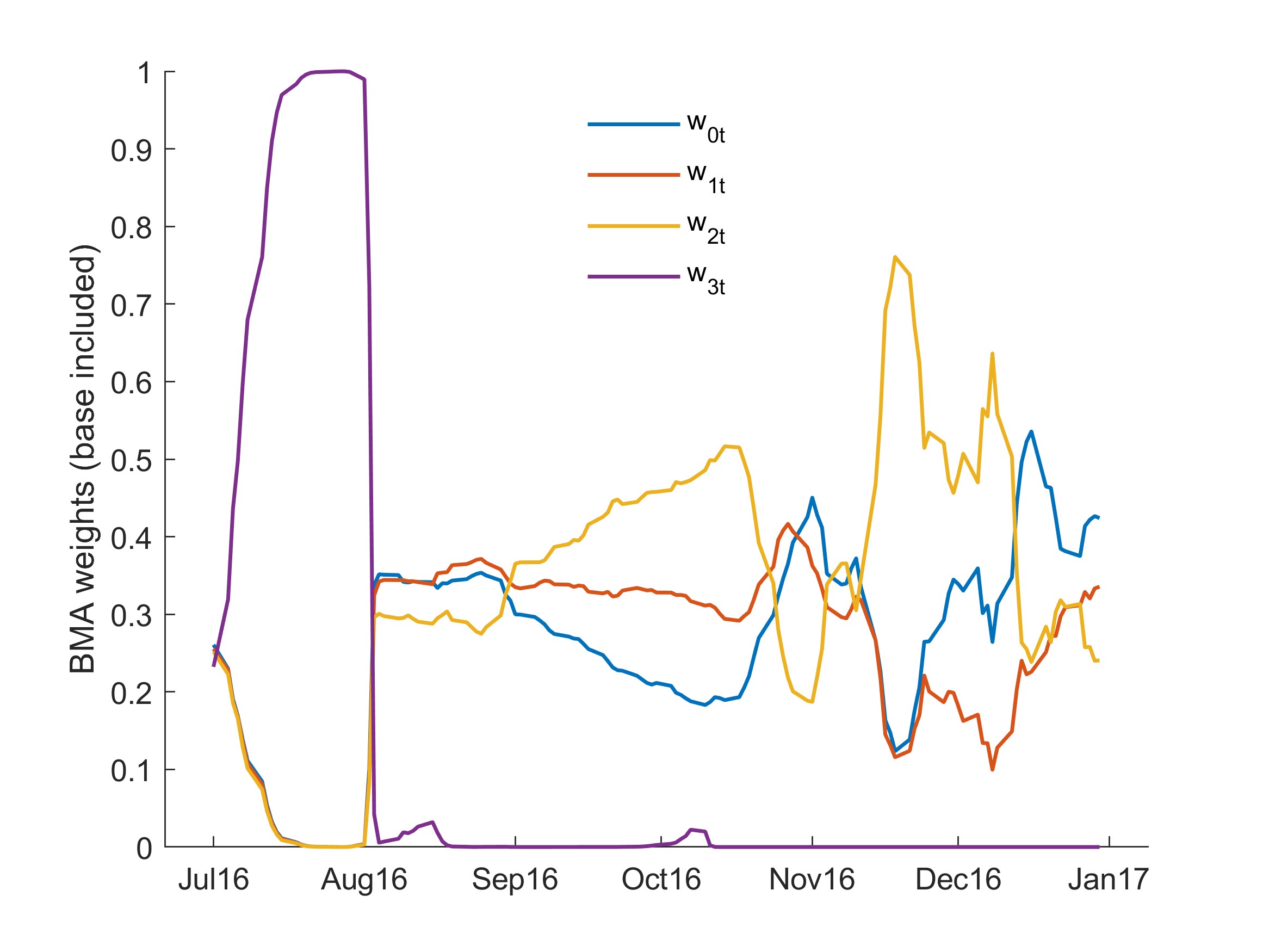}%
  \label{fig:bma_with_base}
}
\caption{Time trajectories of model probabilities:  (a) from traditional BMA using the $J=3$ models; and (b) from the ad-hoc extension to BMAx that includes the baseline model forecasts from the BPS analysis.  }
\end{figure}

The trajectories of BPS weights contrast with those of  BMA model probabilities; the latter are shown in Figure~\ref{fig:bma}. 
As an additional comparison,  Figure~\ref{fig:bma_with_base} shows  BMA analysis extended to included the BPS baseline forecast model as if it were one of the models available to BMA analysis.  The theory of BPS explicitly allows and recommends a baseline, but BMA does not and cannot, since it is defined wholly on the initial model set.  This is the root cause of  the model set completeness issue that bedevils BMA.   Here, the extended analysis reflected in  Figure~\ref{fig:bma_with_base} adds $\mM_0$ to the BMA simply to advantage BMA in the comparison; since this is an ad-hoc extension of BMA, this is denoted as BMAx. 

In this empirical study, BMA effectively settles on an even split of weights between the TVAR models after an initial learning period that showed preference for the locally linear DLM.  Under BMA, posterior model probabilities eventually converge on a single model, so with additional data it is expected that the BMA weights will favor one of the TVAR models. Since BMA scores 1-day ahead forecasts, this is likely to be the simpler $\mM_1$ or $\mM_2.$ In contrast,  
BPS is  open to any one model being favored over time and will not degenerate to any one (wrong) model as samples accrue. This is a theoretical feature of BPS, and an aspect that complements the impact of using a baseline distribution to allow for model set incompleteness.
In this example, BPS   prefers the DLM that BMA discards, unless its predictions are too far from  expectations conditional on predictions from 
the TVAR models. When the preferred model makes forecasts that are too extreme, BPS  falls back to favor the baseline model. Note that although $\mM_1$ and $\mM_2$ receive very little weight in the forecast combination itself, these forecasts are still used to balance the up/down-weighting of all models against the baseline $\mM_0.$ 

In terms of predictive accuracy, some summaries are given in Table~\ref{tableresults}.  In addition to comparing the individual models, BPS, BMA and BMAx, the summary includes  an equally-weighted linear pool of the forecast densities from each of the 3 models (POOL), and an advantaged extension that 
is an equally-weighted linear pool of the models plus the BPS baseline (POOLx).   These are compared on the basis of traditional root mean square error of point forecasts (RMSE) as well as the realized value of the logs of the p.d.f.s of the 5-day forecast distributions (Log Score), each averaged over the last six months of 2016. Evidently, 
BPS outperforms each of the models as well as the two versions of both BMA and POOL.   As with other studies using different BPS model forms~\citep{McAlinnWest2018} and as already noted above, it is no surprise that BMA is less accurate in multi-step forecasting since it inherently scores 1-step ahead accuracy in the prior-posterior model weight updates. 
\begin{table}[t!]
\centering
\begin{tabular}{lll}
Method  & RMSE  & Log Score \\
\hline
BPS     & \bf{1.00} & \bf{1.000}     \\
BMA     & 1.09 & 0.956     \\
BMAx &1.08 &0.956\\
POOL    & 1.05 & 0.965     \\
POOLx &1.06 &0.963 \\
\hline
$\mM_0$ TVAR(1) & 1.09 & 0.946     \\
$\mM_1$ TVAR(2) & 1.10 & 0.946     \\
$\mM_2$ TVAR(5) & 1.11 & 0.945     \\
$\mM_3$ DLM     & 1.15 & 0.886    
\end{tabular}
\caption{Summary measures of 5-day ahead forecast accuracy, normalized to BPS: the root mean square error (RMSE) of point forecasts based on 5-day ahead forecast means, and the log predictive scores (the log of the predictive density for the current day $t$ based on predictions made on day $t-5$) for each model and pooling method.  These metrics are computed each day and averaged over the last six months of 2016.}
\label{tableresults}
\end{table}

\subsection{More on Cross-Model Dependence}
 
The nature of cross-model dependencies is impacted by the forms of the realized forecast model distributions and choices underlying the BPS analysis. The outcome-dependent weighting ability of BPS leads to learning on cross-model dependencies that impact on effective pooling weights in the synthesis.   Elements of this at the BPS level include the time-varying bias vector $\bbeta_t$,  the time-varying cross-model dependence matrix $\bSigma_t$, and the evolving base weights $\q_t.$   Then, perhaps more importantly in applications is the nature of the underlying model set.  A model set that includes a collection of \lq\lq very similar'' models-- similar in terms of generating concordant predictive distributions-- will yield inferences on model dependencies that indicate the strong herding effect.  Models that are more diverse in terms of the predictions they make will, and should, lead to inferences suggestive of weak cross-model dependencies and effective \lq\lq decoupling,'' which can ease interpretation.   

In Section~\ref{sec:BPSFXdiscuss}, 
the model structures are similar but the choice of relatively low discount factors within each model leads to some diversity in model-specific adaptability  to incoming data over time. This is coupled with the focus on 5-day ahead forecasting.  With this forecasting horizon,  shorter-lag TVAR  models show increasing differences relative to  higher-lag TVAR models that can represent momentum effects in FX prices over a few days.  This focus can enhance the ability of BPS to more highly weight longer-lag models in the synthesis pool, while also leading to weaker cross-model dependencies than with less adaptive models and shorter-term forecasting foci. 

\begin{figure}[tb!]
\centering
\includegraphics[clip,width=.5\columnwidth]{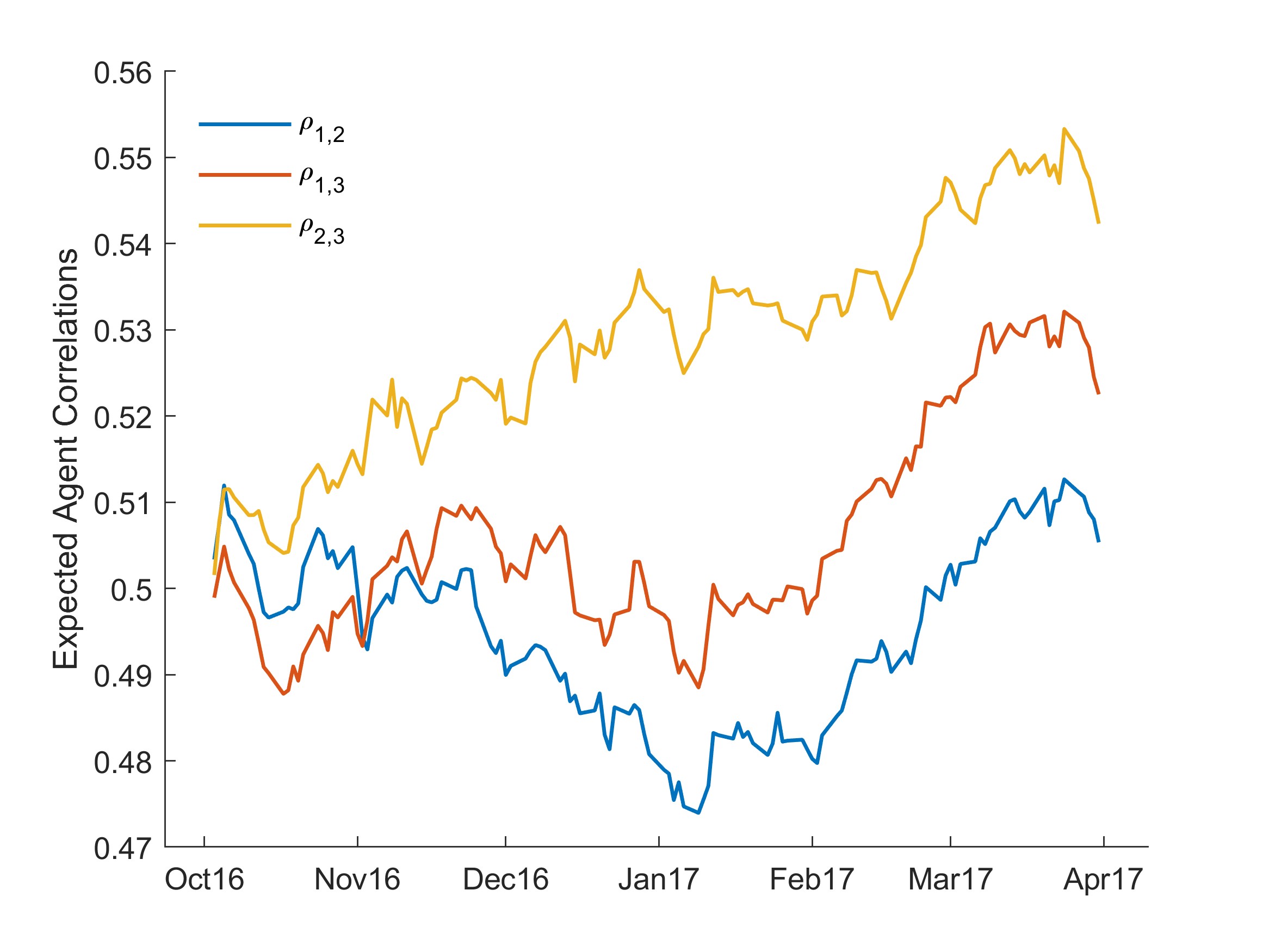}
\caption{Time trajectories of filtered cross-model correlations when using a pool of less time-adaptive models that evidence a stronger herding effect.}
\label{fig:Scorr2}
\end{figure}

To highlight these aspects further, note that repeat analysis focused on 1-day ahead forecasting yields time trajectories of estimated correlations that are positive and much higher-- in the 0.3-0.5 range. This bears  out the reality that these models are very similar in terms of short-term forecasting, but much more distinguished in the BPS analysis based on the longer-term predictive performance.   To further investigate this, consider the roles of (i) differing degrees of adaptability to data in the set of models, based on differing discount factors, and (ii) variants of the choice of synthesis function using the same model structures with different discount factors.  An example using a synthesis function that responds to both consensus and herding effects-- but, critically-- with higher values of the within-model discount factors-- underlies the point estimates of cross-model dependencies shown in Figure~\ref{fig:Scorr2}. The higher, positive correlations here reflect much stronger herding effects due heavily to the constraints within each model to slower adaptation to incoming data enforced by the use of high discount factors for the model-specific state vectors and volatilities.  

\section{Closing Comments}

This paper provides an overview of the BPS framework for forecast model calibration, comparison and combination, and a detailed 
development of specific classes of mixture model-based BPS for forecast density pooling.   While overviewing the general BPS approach and linking to recent developments of various stylized versions of BPS, a main focus here is the subclass of BPS models that yields linear mixture pooling of predictive distributions from each of a set of models.   Discussion details the meanings and implications of outcome-dependent mixture weighting, putting into this foundational BPS setting a number of historical forecast pooling approaches and more recent, important developments in the Bayesian forecasting and econometrics literatures. 
In addition to allowing a decision maker to incorporate beliefs about specific forecasting models, BPS allows practitioners to understand the assumptions of commonly used methods for forecast combination-- including simple pooling with equal (or other) weights on models, traditional Bayesian analysis underlying BMA, and a range of novel and practically relevant extensions that focus on interactions and resulting dependencies across models.  In addition to reflecting model dependencies, a critical dimension of BPS pooling is the explicit, theoretically implied need to admit that \lq\lq all models are wrong,'' with the introduction of a baseline forecast density as a global alternative-- or safe-haven-- to weigh against the forecasts from the chosen set of models.  This theoretically required component addresses long-standing questions in the traditional model averaging and combination literature: the issue of model set incompleteness.     Then, the paper discusses extensions to time series settings, simply by adapting the core BPS theory to allow time-varying parameters underling the supra-Bayesian view of forecast combination defined by the theory of BPS.   The series of examples of theoretically justified forecast density calibration and combination rules emerging from mixture-based BPS, and the detailed example in extension to sequential forecasting in an easily-accessibly FX time series setting, highlight the foundations and methodological opportunities. 
 
Looking ahead, there are several immediate areas of research connections and for potential development. One key point is that, in general, it is not a requirement that the model densities $h_j(\cdot)$ are predictive densities for $y$. These could instead be densities for values related to $y$, and $\mD$ synthesizes this related information using $\alpha(y|\x)$. For example, suppose $y$ is tomorrow afternoon's closing price of the stock for a certain company, and $\mD$ has available density forecasts $h_1(\cdot)$ for that company's quarterly earnings, which will be announced sometime before tomorrow's close, and $h_2(\cdot)$ for a relevant stock index.  The general BPS framework admits such examples. Then, the overall setting of combining forecasts also links to the broader literatures on other approaches, with other desiderata, for predictive combination~\citep[e.g.][]{West1984} and on bringing constraints to predictive models-- whether deterministic or partial constraints on point forecasts or on full forecast distributions~\citep[e.g.][and references therein]{Koop2019,Koop2020,West2023constrainedforecasting}. Further,  the developments here are open to extension to involve additional, model-specific or external information in structuring choices of the central synthesis functions defining BPS.  Some recent developments include ideas to exploit information on historical predictive performance of models~\citep{LavineLindonWest2021avs}, and to explicitly  integrate intended uses of BPS model predictions in resulting decision settings~\citep{TallmanWest2022}, are noted.  The foundational context and theory of BPS has broadened understanding of the scope of subjective Bayesian analysis in the setting of model uncertainty and its roles in prediction, and opened up a number of challenging and interesting directions for future development.

\begin{appendix}


\section{Gibbs Sampler}
\label{sec:suppGibbs}
This section details the Monte Carlo sampling of $(\q_t,\x_t,\bbeta_t,\bSigma_t|y_t)$.  In what follows, the $t$ subscript is omitted in notation, for clarity, with the understanding that sampling takes place at each single point in time after observing $y_t$.

The Gibbs sampler has some complications due to the discrete nature of the mixture synthesis model~\eqref{eq:cross_weights}. This is partly addressed by augmenting with a latent variable $z\in(0:J)$ that denotes the component of the mixture; then
\begin{align*}
\omega_j(\x) &= P(z=j|\q,\x,\bbeta,\bSigma)\\
&=\begin{cases}
1-\sum_{j=1:J}q_j\alpha_j(\x,\bbeta,\bSigma), &\quad j=0,\\
q_j\alpha_j(\x,\bbeta,\bSigma), &\quad j>0,\end{cases}
\end{align*}
where the extended notation now makes explicit that $\alpha_j(\cdot)$ depends on all three parameters. Then, the conditional likelihood is
\begin{equation*}
\alpha(y|\q,\x,z=j,\bbeta,\bSigma)=\begin{cases}
h_0(y), &\quad j=0,\\
\delta_{x_j-\beta_j}(y), &\quad j>0.
\end{cases}\end{equation*}
Further,  $(y\ci\q,\xj,\bbetaj,\bSigma|x_j,z=j,\beta_j)$, so that for $j>0$, $\alpha(y|\q,\x,z=j,\bbeta,\bSigma)=\alpha(y|x_j,z=j,\beta_j)$, while $\alpha(y|\q,\x,z=0,\bbeta,\bSigma)=\alpha(y|z=0)$. This is partially evident from the construction of the directed graph of the model,  and the associated conditional independence graph: 

$$\vcenter{\xymatrix{
&\x\ar[d]\ar[rd]  &   &  &\x\ar@{-}[dl]\ar@{-}[dr]\ar@{-}[ddl]\ar@/_1pc/@{-}[dd]\ar@{-}[d]&\\
\q\ar[r]&z\ar[r]&y  &        \q\ar@{-}[rd]\ar@{-}[d]\ar@{-}[r]&z\ar@{-}[r]&y\\
\bSigma\ar[r]\ar[ur]&\bbeta\ar[u]\ar[ur]&  &
\bSigma\ar@{-}[ur]&\bbeta\ar@{-}[l]\ar@{-}[u]\ar@{-}[ur]&
}}$$
The joint density is then
\begin{equation*}
p(\q,\x,y,z=j,\bbeta,\bSigma) = \alpha(y|x_j,z=j,\beta_j) \\
P(z=j|\q,\x,\bbeta,\bSigma)  p(\q)h(\x)p(\bbeta,\bSigma),
\end{equation*}
where $h(\x)$ is the product of model forecast densities.


\subsection{Sampling \texorpdfstring{$z$}{z}}

Analysis samples $(z|\q,y,\bbeta,\bSigma)$ (with $\x$ marginalized out), and then $(\x|q,y,z=j,\bbeta,\bSigma)$. 
Start with 
$$
P(z=j|\q,y,\bbeta,\bSigma) \propto\\P(z=j|\q,\bbeta,\bSigma) p(y|\q,z=j,\bbeta,\bSigma).
$$
The first term here is
\begin{align*}
P(z=j|\q,\bbeta,\bSigma)&=\int P(z=j|\q,\x,\bbeta,\bSigma)h(\x)d\x\\
&=\begin{cases}
1-\sum_{j=1:J}q_j\int\alpha_j(\x,\bbeta,\bSigma)h(\x)d\x,&j=0,\\
q_j\int\alpha_j(\x,\bbeta,\bSigma)h(\x)d\x,&j>0.
\end{cases}.
\end{align*}
The integrals here are evaluated via direct Monte Carlo integration based on samples from the $h_j(\cdot).$  
The second term has closed form
\begin{align}
p(y|\q,z=j,\bbeta,\bSigma)&=\int\alpha(y|\q,\x,z=j,\bbeta,\bSigma)h(\x)d\x\\
&=\begin{cases}
h_0(y),&j=0,\\
h_j(y+\beta_j,)&j>0.
\end{cases}
\end{align}
Normalizing the resulting product of these two terms gives the desired probabilities $P(z=j|\q,y,\bbeta,\bSigma)$, and 
these are used to resample $z.$


\subsection{Sampling \texorpdfstring{$\x$}{x}}
The full conditional density for $(\x|q,y,z=j,\bbeta,\bSigma)$ breaks down into two cases depending on the value of $z$. When $z=j>0$, the likelihood for $y$ depends on $x_j$:
$$
p(\x|\q,y,z=j,\bbeta,\bSigma)\propto\\\delta_{x_j-\beta_j}(y)  \alpha_j(\x,\bbeta,\bSigma)  h(\x).
$$
In this case,   $x_j=y+\beta_j$, and the remainder of the $\x$ vector is filled in using 
rejection sampling with acceptance probability $\alpha_j(y+\beta_j,\xj,\bbeta,\bSigma)$.
When $z=0$,
\begin{equation*}
p(\x|y,z=0,\bbeta,\bSigma)\propto \omega_0(\x,\bbeta,\bSigma)  h(\x).
\end{equation*}
Rejection sampling is again used, this time with acceptance probability $$\omega_0(\x,\bbeta,\bSigma)=1-\sum_{j=1:J}q_j\alpha_j(\x,\bbeta,\bSigma).$$


\subsection{Sampling \texorpdfstring{$(\bbeta,\bSigma)$}{(beta, Sigma)}}
The full conditional density for $(\bbeta,\bSigma)$ is  
\begin{equation*}
p(\bbeta,\bSigma|\q,\x,y,z=j)\propto\\ \alpha(y|\q,\x,z=j,\bbeta,\bSigma)P(z=j|\x,\bbeta,\bSigma)  p(\bbeta,\bSigma).
\end{equation*}
Split the sampler into cases $z=0$ and $z>0$.
When $z=0$, rejection sampling is straightforward.
$\alpha(y|\q,\x,z=j,\bbeta,\bSigma)=h_0(y)$, so $(\bbeta,\bSigma)$ may be sampled from their prior and accepted with probability $$\omega_0(\x)=1-\sum_{j=1:J}q_j\alpha_j(\x,\bbeta,\bSigma).$$
When $z=j>0$, $\beta_j$ is defined as $x_j-y$.
Then $\bSigma$ is sampled from its inverse-Wishart prior, $\beta_j=x_j-y$, and the remainder of $\beta$ is sampled from its normal prior distribution conditional on $\bSigma$ and $\beta_j$.
The sample is accepted with probability $\alpha_j(\x,\bbeta,\bSigma)$.


\subsection{Sampling \texorpdfstring{$\q$}{q}}

The full conditional density for $(\q|\x,y,z=j,\bbeta,\bSigma)$ also breaks down into two cases. The only relevant terms are $P(z=j|\q,\x,\bbeta,\bSigma)$ and $p(\q)$. When $z=j>0$,
\begin{equation*}
p(\q|\x,y,z,\bbeta,\bSigma)\;\propto\; q_j  p(\q).
\end{equation*}
If $p(\q)$ is a Dirichlet density with parameters $(u_1,\dots,u_J)$, this allows for a conjugate update and exact sampling with $u_j\rightarrow u_j+1$. When $z=0$,
\begin{equation*}
p(\q|\x,y,z,\bbeta,\bSigma)\;\propto\; \omega_0(\q,\x,\bbeta,\bSigma)  p(\q)
\end{equation*}
where
\begin{equation*}
\omega_0(\q,\x,\bbeta,\bSigma)=1-\sum_{j=1:J}q_j\alpha_j(\x,\bbeta,\bSigma).
\end{equation*}
Rejection sampling is again used with acceptance probability $\omega_0(\q,\x,\bbeta,\bSigma)$.


\section{Variational Bayes}
\label{sec:vb}

In the time series context with sequential forecasts,  assumed parametric  forms of the priors for $(\bbeta_t,\bSigma_t$) and $\q_t$ are adopted at each time step $t$. Since posteriors at  the previous time are represented in terms of Monte Carlo samples, the constraint to specific parametric forms for the current time are imposed  using variational Bayes (VB).  This identifies parameters of the approximating parametric forms that minimize the K\"{u}llback-Leibler (KL) divergence of the approximating distribution from that of the posterior samples. The underlying conceptual basis, and resulting methodology,  is similar to that of~\cite{GruberWest2016BA}, in which the authors fit a normal-inverse-gamma distribution to posterior samples at each time point by minimizing the KL divergence of the approximating parametric form from the distribution represented by the Monte Carlo sample. This involves a combination of 
analytic solutions for some of the parameters and a simple numerical optimization for others, as follows. 
 
\subsection{Normal Inverse Wishart Approximation}
Write the joint NIW distribution such that $(\bbeta|\bSigma)\sim N(\bb,c\bSigma)$ and $\bSigma\sim IW(n,\S)$. Using this notation, the optimal parameters are given by
\begin{enumerate}
\item $\bb=E[\bSigma^{-1}]^{-1}E[\bSigma^{-1}\bbeta]$;
\item $c=E[(\bbeta-\bb)'\bSigma^{-1}(\bbeta-\bb)]/J$;
\item $n$ satisfies
\begin{equation*}
E[\log(|\bSigma|)]+\log(|E[\bSigma^{-1}]|)-J\log((n+J-1)/2)\\
+\sum_{j=1:J}\psi((n+j-1)/2)=0    
\end{equation*}
where $\psi(\cdot)$ denotes the digamma function $\psi(x)=\Gamma'(x)/\Gamma(x)$;
\item $\S=E[\bSigma^{-1}]^{-1}(n+J-1)/n$.
\end{enumerate}
The expectations here are computed from the Monte Carlo sample. Note that $\bb$ and $c$ are directly evaluated, a simple Newton-Raphson optimisation generates $n$ and then $\S$ is directly computed. This setting is a complete parallel to that in~\cite{GruberWest2016BA} with the simple extension of normal, inverse gamma distributions there to NIW distributions here. 


\subsection{Dirichlet Approximation}
For a $Dir(u_1,\dots,u_J)$ approximation to the Monte Carlo posterior samples of the $J$-vector $\q$ on the simplex, the parameters $u_1,\dots,u_J$ satisfy
\begin{equation*}
\psi(\sum_{j=1:J}u_j)+\psi(u_i)-\sum_{j=1:J}E[\log(q_i)]=0
\end{equation*}
where $\psi(\cdot)$ again denotes the digamma function $\psi(x)=\Gamma'(x)/\Gamma(x)$. An analytical solution is not available, but an approximate solution is again trivially implemented using a multivariate Newton-Raphson analysis, solving the above equation within an arbitrary tolerance.


\section{Application Details}
\label{sec:app_details}

This section contains additional details for the application in Section~\ref{sec:application}.

\subsection{Model Specification}
The example in Section~\ref{sec:application} combines predictive densities from 4 pure time series models with time-varying parameters.
All models are initialized at year-end 2015 and trained for the first half of 2016 before providing daily 5-step ahead forecasts for the second half of 2016.
The models are trained from January 1, 2016 through June 24, 2016 (130 training observations) before producing the first 5-step ahead forecast for July 1, 2016.

 A full description of the baseline and model densities requires detailing of initial priors and discount factors on model parameters.
The specifications summarized here   use the standard notation of \cite{West1997} and \cite{PradoFerreiraWest2021} for each of these univariate dynamic linear models. The standard notation uses $\btheta_t$ for the model state vector and $v_t$ for the variance of observations around the dynamic linear regression over time $t.$  Each model has an initial normal prior $N(\m_0,\mathbf{C}_0)$ on $\btheta_0$ and an inverse-gamma $IG(n_0/2,d_0/2)$ (with harmonic mean $s_0$) on $v_0$.  These were chosen to reflect relatively vague initial priors for the example analysis. 
In each model, evolution variances for the coupled random-walk evolutions of $\btheta_t$ and $v_t^{-1}$ are specified through the use of two discount factors, one for the state vector and one for the residual variance.

The TVAR models differ only in the chosen AR lag. The priors in each  have the following features: 
prior mean of the lag$-1$ AR parameter is 0.97, that for higher-order AR parameters is 0.  The initial prior mean for the intercept in the auto-regression of each model is set so that   $E[y_1]=y_0$, the last daily value of the time series before the start of the data analysis time period.
The initial variance matrix for the model state vector $\mathbf{C}_0$ is diagonal with entries $10^{-4}$. 
The initial inverse-gamma prior for $v_0$ is defined by $n_0=10$ and $s_0=0.01$.
The discount factors for both the latent state vector and the residual variance are set to 0.95.  

The locally linear DLM has initial prior as follows.  The prior mean for the local level (intercept) at $t=1$ is $y_0$, and that for the gradient from $t=0$ to $t=1$ is zero.  The prior variance matrix  $\mathbf{C}_0 = \textrm{diag}[10^{-4}, 10^{-5}].$ 
The inverse-gamma distribution prior for $v_0$ is defined via $n_0=10$ and $d_0=0.001$.
The state and residual variance discount factors are set relatively low at 0.9 to allow faster adaptation to daily variation in FX series. 

At each time point, predictive densities are sampled using standard methodology, projecting the model-based forecasts to 5-days ahead in terms of a Monte Carlo sample for each model. Then, a scale- and location- shifted Student$-t$ distribution is fitted to each of the resulting Monte Carlo  samples; this  uses the Mathworks {\verb|fitdist|} function.  These resulting T distributions are taken to define the inputs to the BPS, BMA and POOL analyses.  

\subsection{Synthesis Function}

For clarity, again drop the $t$ subscript and denote weight functions as $\omega_j(\x)$ and $\alpha_j(\x)$, with the implicit understanding that they also depend on $\bmu$ and $\bSigma$; recall that $\bmu$ is defined as the point forecast of the baseline density $h_0(y)$ with offsets given by a bias vector $\bbeta$, while $\bSigma$ represents cross-model dependencies.

The synthesis function is as defined in \eqno{cross_weights_time} with the additional specification
\begin{subequations}
\begin{equation*}
\omega_j(\x_t) = q_j\alpha_j(\x)
\end{equation*}
for $j>0$, with
\begin{equation*}
    \omega_0(\x) = 1 - \sum_{\jj}\omega_j(\x).
\end{equation*}
\end{subequations}
For $j>0$, set
$$\alpha_j(\x) = \exp\{-e_j^2/(2\nu_j)\}$$ 
where $e_j=x_j-\mu_j-\bgamma_j'(\xj-\muj)$, and $\nu_j$ and $\bgamma_j$ represent the conditional variance and regression vector implied by $\bSigma$.
This construction accounts for model consensus by discounting $x_j$ far from the conditional expectation. 

\subsection{BPS Priors and Discount Factors\label{sec:Appdiscounts} }

Consider first the NIW prior specified at $t=0$ for $(\bbeta,\bSigma)$.
As the data do not directly inform on $(\bbeta,\bSigma)$, $\mD$'s subjective prior at time $t=0$ can be especially important.
The reported analysis adopts an inverse-Wishart prior with $n_0=15$ degrees of freedom and point estimate $\S$.
$\S$ is diagonal with elements equal to $s_{0,1}$, the (squared) scale parameter from the baseline density for $y_1$, reflecting an uninformed prior.
The bias vector $\bbeta$ has conditionally normal distribution $N(\bb, r\bSigma)$.
The time $t=0$ conditional normal prior for $\bbeta$ is likewise centered at zero, so that $b_j=0$ initially for all $j$.
Set $r_0=1$, implying a prior variance for the latent states $\x$ given $\bSigma$ alone that is double that of the conditional variance given both $\bSigma$ and $\bbeta$.
The time Dirichlet prior for the base synthesis weights $\q_0$ is taken as the uniform Dirichlet. 

The time $t$ posteriors for $(\bbeta_t,\bSigma_t)$ and $\q_t$ (having observed $y_t$) are denoted $(\bbeta_t, \bSigma_t|y_{1:t})\sim NIW(\bb_t, c_t, n_t, \S_t)$ and
$(q_t|y_{1:t})\sim Dir(\u_t).$
With respective discount factors $\delta_\beta$, $\delta_\Sigma$, and $\delta_q$, the time $t$ priors for $(\bbeta_{t+1},\bSigma_{t+1})$ and $\q_{t+1}$ are written
$$(\bbeta_{t+1}, \bSigma_{t+1}|y_{1:t})\sim NIW(\bb_t, c_t/\delta_\beta, \delta_\Sigma n_t, \S_t)$$
and
$$(q_{t+1}|y_{1:t})\sim Dir(\delta_q\u_t).$$
That is, $\delta_\beta$ slightly increases the scalar $c_t$ on the conditional variance from $t$ to $t+1$, while $\delta_\Sigma$ slightly decreases in the degrees of freedom in the IW distribution.
In the Dirichlet, scaling the parameters down maintains the same point estimates while increasing uncertainty.
The dynamic model discount factors $\delta_\cdot<1$ are chosen to indicate the expectation of generally stable patterns of bias and cross-model dependencies over time. 
The discount factor controlling the time evolution of $\bSigma_t$ in section~\ref{sec:application} is 0.98, and those related to the time evolution of $\bbeta_t$ and the Dirichlet time evolution of $\q_t$ are 0.97.
 
\end{appendix}



\bibliographystyle{elsarticle-harv}
\bibliography{JohnsonWest}       

\end{document}